**Analytic and Numerical Models of Oxygen and Nutrient Diffusion, Metabolism Dynamics, and Architecture Optimization in Three-Dimensional Tissue Constructs with Applications and Insights in Cerebral Organoids**

Richard J. McMurtrey[1,2]

[1] Institute of Neural Regeneration & Tissue Engineering, Highland, UT 84003, United States, Email: *richard.mcmurtrey@neuralregeneration.org*

[2] Institute of Biomedical Engineering, Department of Engineering Science, Old Road Campus Research Building, University of Oxford, Oxford, OX3 7DQ, United Kingdom, Email: *richard.mcmurtrey@eng.oxon.org*

---

## Abstract

Diffusion models are important in tissue engineering as they enable an understanding of molecular delivery to cells in tissue constructs. As three-dimensional (3D) tissue constructs become larger, more intricate, and more clinically applicable, it will be essential to understand internal dynamics and signaling molecule concentrations throughout the tissue. Diffusion characteristics present a significant limitation in many engineered tissues, particularly for avascular tissues and for cells whose viability, differentiation, or function are affected by concentrations of oxygen and nutrients. This paper seeks to provide novel analytic solutions for certain cases of steady-state and non-steady-state diffusion and metabolism in basic 3D construct designs (planar, cylindrical, and spherical forms), solutions that would otherwise require mathematical approximations achieved through numerical methods. This model is applied to cerebral organoids, where it is shown that limitations in diffusion and organoid size can be partially overcome by localizing metabolically-active cells to an outer layer in a sphere, a regionalization process that is known to occur through neuroglial precursor migration both in organoids and in early brain development. The given prototypical solutions include a review of metabolic information for many cell types and can be broadly applied to many forms of tissue constructs. This work enables researchers to model oxygen and nutrient delivery to cells, predict cell viability, design constructs with improved diffusion capabilities, and accurately control molecular concentrations in tissue constructs that may be used in studying models of development and disease or for conditioning cells to enhance survival after insults like ischemia or implantation into the body, thereby providing a framework for better understanding and exploring the characteristics of engineered tissue constructs.

---

# Introduction

An understanding of diffusion limitations in tissues is essential for studying not only cell survival but also many forms of cellular functions. In particular, oxygen and nutrients can be limited in tissue cultures, as these must diffuse from gas and liquid phases into a solid phase composed of individual cells, cell clusters, extracellular matrix, hydrogels, or other materials in order to reach the cells. Gas and nutrient levels in tissues have begun to be appreciated for their significant effects on stem cell proliferation, differentiation, and overall function, as mediated through numerous pathways, with oxygen affecting stem cell states[1-18], gene transcription[19-22], neurotransmitter metabolism[23-26], and cell viability[11,27-30]. In addition, other key nutrients, such as glucose, lipids, amino acids, cell signaling molecules, and growth factors, must also diffuse through cells and tissues, and even small variations in their concentrations can affect cell differentiation, development, and function. Therefore a detailed understanding of the internal dynamics of oxygen and nutrient diffusion and metabolism is essential in studying cell and tissue functions.

Recent work has demonstrated distinct advantages of three-dimensional (3D) cultures for many types of tissue, particularly for replicating architecture of neural tissue[31-34]. However, 3D tissue constructs quickly acquire significant diffusion limitations as the size and cell density is increased, and diffusion limitations are one of the primary prohibitive factors in scaling up large 3D tissue models[35-36]. The ability to model diffusion and availability of nutrients and gasses to the cells is thus an important consideration in the design of tissue constructs, and with the advent of organoid cultures and more complex 3D tissue models, modeling and analysis of nutrient delivery to cells becomes ever more important. Diffusion models, however, require an understanding of complex differential equations, and prior models of diffusion have only begun to explore applications to tissue constructs, focusing on numerical solutions that require specialized software and programming capabilities. Moreover, the specific source code and formulations are generally not made available, and even when source code is available it applies only to a particular system and set of conditions. General methods for numerically solving difficult differential equations were developed by Euler in the 18th century and Runge and Kutta in the 19th century, and many more advanced methods have been and still are being developed. However, equations and models which are reducible to closed-form solutions are extremely useful in their ease of application as well as elegant in their forms, yet investigations have not yet been made into complete analytic models and solutions that are broadly applicable to 3D tissue constructs.

This paper therefore first seeks to provide novel analytic or closed-form solutions for certain mass transfer models in order to enable any researcher to estimate molecular dynamics and diffusion characteristics for a particular tissue construct. Metabolic characteristics of a variety of cell types are presented to evaluate viable diffusion and metabolism of oxygen and nutrients in a variety of tissue construct scenarios, and previously undescribed analytic models for a variety of tissue construct designs are shown to apply to experimental characteristics in single- and multi-dimensional models. The necessary components for prototypical analytical models are described, and although this work demonstrates the derivations and solutions for several unique applications of challenging differential equations, only a working knowledge of algebra is ultimately needed to utilize the final formulas. These models are then applied to cerebral organoids to obtain a better understanding of their characteristics and functions. While this paper focuses primarily on applications to neural tissue engineering, these approaches and solutions also generally apply to any type of tissue, organ, or implant application and to any type of diffusant molecule, including modeling and analysis of cell function and viability in a variety of 3D tissue constructs.

| Nomenclature Summary | | | |
|---|---|---|---|
| $a$ | Coefficient determined by boundary conditions | $n$ | Index term in sigma summation series |
| $A$ | Surface area | $\Omega$ | Proportion of cells in outer spherical shell |
| $B$ | Biot coefficient | $P$ | Partial pressure of gas |
| $b$ | Coefficient determined by boundary conditions | $\pi$ | 3.14159…. |
| $C$ | Concentration | $\varphi$ | Metabolic consumption rate for tissue construct |
| $C_o$ | Initial concentration outside tissue construct | $\varphi_1$ | Internal shell $\varphi$ for modified architecture |
| $\bar{C}$ | Average concentration in tissue construct | $\varphi_2$ | External shell $\varphi$ for modified architecture |
| $C_i$ | Initial concentration inside tissue construct | $R$ | Outer radius of a radial tissue construct |
| $C_{critical}$ | Critical concentration | $R_{max}$ | Maximal radius of a radial tissue construct |
| $C_{media}$ | Media concentration | $R_{max_t}$ | Maximal radius as a function of time |
| $C_{net}$ | Net concentration in construct | $R_{opt}$ | Modified $R_{max}$ based on alterations in architecture |
| $C_s$ | Concentration at outer rim of inner spherical shell | $R_g$ | Modified $R_{max}$ of inner spherical shell |
| $D$ | Diffusion coefficient | $r$ | Radial distance or spatial position |
| $d$ | Ordinary Derivative | $\rho$ | Density of cells in tissue construct |
| $\partial$ | Partial Derivative | $\rho_1$ | Internal shell $\rho$ for modified architecture |
| erf | Error function | $\rho_2$ | External shell $\rho$ for modified architecture |
| erfc | Complementary error function | s | Spatial dimension of system |
| $e$ | 2.71828…. | $\Sigma$ | Summation series in sigma notation |
| $\varepsilon$ | Eigenvalue | $t$ | Time |
| $F$ | Isolated function of distance | $t_{consum}$ | Time to consumption of nutrient |
| $G$ | Isolated function of time | $t_{s.s.}$ | Time to steady state |
| $g$ | Coefficient determined by boundary conditions | $T$ | Thickness of linear tissue construct |
| $\gamma$ | Constant in separation of variables method | $T_{max}$ | Maximal thickness of linear tissue construct |
| $H$ | Henry's constant | $T_{max_t}$ | Maximal thickness as a function of time |
| $h$ | Mass transfer coefficient | $\vartheta$ | Placeholder for integrating $\mu$ |
| $J$ | Flux | $\Upsilon$ | Percent radius between center and surface |
| $J_q()$ | Bessel function of the first kind and order $q$ | $V$ | Volume |
| $l$ | Length through liquid phase to construct surface | $V_c$ | Volume of tissue construct |
| $\lambda_n$ | Eigenvalues | $V_m$ | Volume of media around tissue construct |
| $M$ | Metabolic profile function | $x$ | Spatial position or linear depth into construct |
| $m$ | Metabolic consumption rate per cell | $\xi$ | Position acting as diffusant source |
| $\mu$ | Defined as $\frac{x}{\sqrt{4Dt}}$ | $y$ | Spatial position in Cartesian coordinates |
| $N$ | Amount of diffusant in system | $z$ | Spatial position in Cartesian coordinates |

# Experimental Methods & Materials

Cerebral organoids were cultured for growth analysis according to protocols described in detail by Lancaster and Knoblich[37]. In this case, feeder-independent human induced pluripotent stem cells (hiPSCs) derived from normal fibroblasts were maintained, dissociated, and seeded into embryoid bodies at 9,000 cells per embryoid body in hESC media, marking day 0 of organoid growth and the beginning of germ layer differentiation. When embryoid bodies reached 500-600 μm (typically day 6), neural induction was begun to promote neural ectoderm and neurosphere formation. After 5 days of induction, neurospheres were transferred to 30 μl Matrigel droplets and cultured in cerebral organoid differentiation media without vitamin A for the first four days and then with vitamin A and orbital shaking for all time thereafter. The primary differences from published protocols were that the iPSCs were cultured and passaged in Essential 8 media (Life Technologies), and the hESC media (with low FGF-2 and Y-27632 on days 0-4 and without on days 4-6) was instead composed of 390 ml DMEM/F12, 100 ml KOSR, 5 ml MEM-NEAA, 5 ml Glutamax, and 3.6 μl of β-mercaptoethanol. Neurospheres were embedded in Matrigel on day 11 and maintained in 60 mm dishes with 5 ml of cerebral organoid differentiation media and approximately five organoids in each dish. The sizes of the organoids were measured and recorded over time; for imperfectly shaped spheroidal structures, the shortest radial component between the deepest point and the surface was measured.

Images of whole organoids were taken on an inverted phase-contrast light microscope, and fluorescent imaging of sections was performed on a Nikon A1 confocal microscope with excitation wavelengths of 405 nm and 488 nm. Analysis of organoid cell density was determined by washing the organoids with phosphate buffered saline (PBS) without calcium or magnesium, dissociating ten separate organoids with 1000 μl of Accutase at 37°C and counting cells with a hemocytomer. The total number of dissociated cells from each organoid was divided by the estimated volume of each organoid, measured as the unionized volume of 3D volume models over the scaled microscopic images, and reported as average cell density and standard deviation, values which were used in the calculation of metabolic parameters and diffusion characteristics in spherical models of organoids. For sections, organoids were fixed in 4% (w/v) paraformaldehyde, washed with 1X PBS, suspended in 30% sucrose (w/v), then embedded in standard optimal cutting temperature (OCT) compound, frozen on dry ice, and sectioned into 25 μm thick sections that were placed on superfrost plus slides. Sections were then blocked with a solution of 1X PBS, 3% (w/v) bovine serum albumin (BSA) and 0.1% tween-20 for 2 hours, incubated with anti-beta-III-tubulin TUJ1 primary antibody (Covance) at 1:750 dilution at 4°C overnight, and washed three times with a wash solution of 1X PBS and 0.1% tween-20. Samples were then incubated with Alexa-Fluor 488 secondary antibody (Life Technologies) at 1:1000 dilution and Hoechst stain (1:1000) for 2 hours, then washed three times with wash solution, and samples were covered with mounting media and coverslips. Mathematical models were developed and applied as described in the text.

# Modeling and Results

*1.1 Quasi-Steady-State Solution in Linear 1D Cases*

Three-dimensional shapes have often been modeled with one-dimensional solutions in applications such as thermodynamics and diffusion, including in common configurations of planar, cylindrical, and spherical constructs[38-42], and numerically-solved one-dimensional diffusion models have been shown to be an accurate and valid simplification for many types of three-dimensional tissue constructs[43-44]. One-dimensional analytic solutions are also applicable to diffusion in tissue constructs, particularly in constructs where the majority of diffusion occurs in a uniform direction (designated as distance $x$, see Figure 1) or where diffusion occurs through a primary surface such as the top layer of a planar or slab-like construct. One-dimensional solutions are also valid when diffusion occurs through the outer surface of a cylinder along a radial axis in cylindrical coordinates or through the outer surface of a spherical construct in spherical coordinates along the radial axis (designated as radial distance $r$, see Figure 1). Solutions with circular or spherical symmetry are producible by altering the coordinate frame and adjusting boundary conditions appropriately for the diffusion system, thereby accurately depicting diffusion and metabolism for specific forms of one-, two-, or three-dimensional constructs.

Diffusion can be modeled using physical laws originally described by Adolf Fick, depicting a changing flux over distance and conservation of mass over distance and time. Fick had a passion for mathematics and physics, and although he chose to become a physician, he continually sought to apply the concepts of mathematics to the study of the human body. Fick's model is based on similar forms of equations for transfer of heat and energy, as described by Joseph Fourier and Isaac Newton respectively, being derived from the conservation of mass within a defined region, wherein the change in concentration ($C$) per time ($t$) is equal to the change in flux ($J$) over distance ($x$), written as $\frac{\partial C}{\partial t} + \frac{\partial J}{\partial x} = 0$. By substituting Fick's first law for flux $J = -D\frac{\partial C}{\partial x}$ into the conservation of mass equation (where $D$ is diffusivity), the classic form of the diffusion equation is obtained:

$$\frac{\partial C}{\partial t} = D\frac{\partial^2 C}{\partial x^2} \quad (1)$$

In order to describe exact solutions to this equation, the one-dimensional case of distance ($x$) into a tissue construct is first considered here (Figure 1). The solutions to Eq. (1) can be described in steady-state and non-steady-state, and can also be given over the infinite, semi-infinite, or finite spatial domains. By maximizing the tissue construct thickness such that complete consumption of nutrient occurs at the deepest point in the construct, certain difficulties in the mathematics can be surmounted. A useful quasi-steady state solution (where the change in concentration of a nutrient in time is constant) can be found for a tissue construct that has characteristics of both diffusion and metabolism by first adding the change in concentration over time to a metabolic component called $\varphi$, representing the consumption rate of the nutrient by the cells in culture (in units of $\frac{mol}{L \cdot s}$ herein). By then letting the diffusional component reach quasi-steady-state $\left(\frac{\partial C}{\partial t} + \varphi = \varphi\right)$, the equation becomes:

$$\varphi = D\frac{\partial^2 C}{\partial x^2} \quad (2)$$

It is reasonable to treat $\varphi$ as a constant (zero order kinetics) when cells are no longer dividing and when they are in a stable metabolic state, although in many conditions there are often phases of growth, differentiation, and consumption that may fluctuate over time. By defining $C_o$ as the concentration of the nutrient in the media at the surface of the hydrogel, defining the surface of the hydrogel as $x=0$, and defining $T$ as the length through the hydrogel from the surface to the floor of the construct (Figure 1), the maximal thickness of tissue construct can be solved by integrating twice. Certain boundary conditions are applied in this process, depending on the construct design. With an impermeable floor at the base of the construct ($x = T$), the change in concentration ceases, given as $\frac{\partial C(x=T,t)}{\partial x} = 0$. Furthermore, with the assumption that the concentration at the surface of the construct remains constant, given by $C(x = 0, t) = C_o$, the integration of equation (2) becomes:

$$C(x) = \frac{\varphi x^2}{2D} - \frac{\varphi T x}{D} + C_o \quad (3)$$

Substituting $C(x) = 0$ at $x = T$ represents the distance at which the nutrient is fully consumed, which then further defines the maximum thickness of diffusion in a cellularized hydrogel:

$$T_{max} = \sqrt{\frac{2C_o D}{\varphi}} \quad (4)$$

If diffusion is allowed from both below and above the construct (such as with planar constructs in a suspended well in the media rather than seated on the floor[34]), the $T_{max}$ is simply doubled.

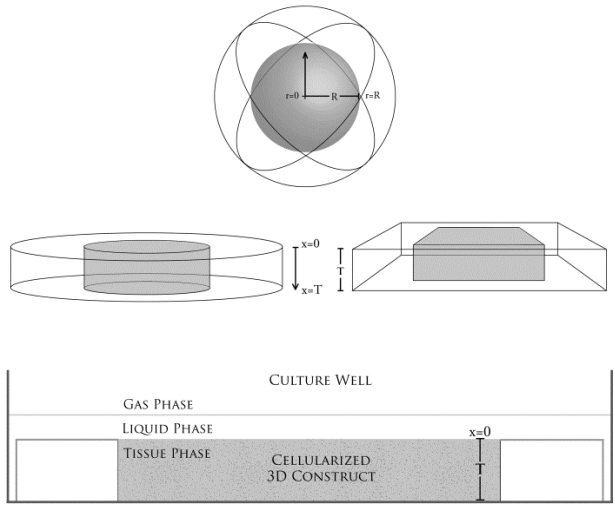

Figure 1: Views of 3-dimensional tissue constructs in a culture well. The thickness of the planar construct is $T$ and the diameter of the sphere is $2R$.

### 1.2 Quasi-Steady-State Solution in Radial Cases

Radially symmetric constructs may also be modeled with one-dimensional solutions by substituting spherical or cylindrical coordinates into the diffusion equation. In the spherical case, just as in aforementioned conditions, it is assumed that density and diffusion coefficients will be constant throughout the construct and that the dimensions of the sphere are constant without swelling or shrinking. By substituting spherical coordinates into the three-dimensional diffusion equation $\frac{\partial C}{\partial t} = D\Delta C$ (where $\Delta$ is the LaPlacian operator, or the divergence of the gradient of $C$), and by implementing azimuthal and polar symmetry around the sphere (meaning the change in concentration with respect to $\theta$ or $\phi$ is zero, as would be expected in an ideal sphere), diffusion in a sphere of radius $R$ along the radial axis $r$ can be written as follows[38,42,45]:

$$\frac{\partial C}{\partial t} = \frac{1}{r^2}\frac{\partial}{\partial r}\left(r^2 D \frac{\partial C}{\partial r}\right) \quad (5)$$

In fact, a simpler way of writing this equation for any dimensionality $s$ is[46]:

$$\frac{\partial C}{\partial t} = \frac{1}{r^{s-1}}\frac{\partial}{\partial r}\left(r^{s-1} D \frac{\partial C}{\partial r}\right) \quad (6)$$

where s=1 describes the one-dimensional axis, s=2 describes a cylinder where diffusion at the end surfaces is negligible and occurs primarily along a circular radius, and s=3 describes a sphere. Assuming a constant nutrient consumption rate and constant isotropic diffusivity sets up the spherical equation similar to Eq. (2):

$$\varphi = \frac{D}{r^2}\frac{\partial}{\partial r}\left(r^2\frac{\partial C}{\partial r}\right) \quad (7)$$

Integrating twice produces:

$$C(r) = \frac{\varphi r^2}{6D} - \frac{c_1}{r} + c_2 \quad (8)$$

Boundary conditions are produced by the condition of spherical symmetry, $\frac{\partial C(r=0)}{\partial r} = 0$, giving $c_1 = 0$, and from the initial concentration at the outer edge of the sphere (at point $R$), the concentration $C(r = R, t) = C_o$. Then $c_2$ can be solved as $c_2 = C_o - \frac{\varphi R^2}{6D}$, giving a general solution for concentration along $r$ in the construct:

$$C(r) = \frac{\varphi}{6D}(r^2 - R^2) + C_o \quad (9)$$

The maximal depth occurs when $C = 0$ at $r = 0$, which can then be solved to find maximal $R$ for a given $\varphi$:

$$R_{max} = \sqrt{\frac{6C_oD}{\varphi}} \quad (10)$$

The $R_{max}$ thus represents the maximal diffusion distance from a cell to a surface source of nutrient, and this has been used, for example, as the maximal viable radius in numerical models of spherical tumor growth[47-49]. By the same methods, the general cylindrical solution, where diffusion occurs through the surface curvature and not at the ends, is of the form:

$$C(r) = \frac{\varphi r^2}{4D} - c_1\ln r + c_2 \quad (11)$$

Under the described boundary conditions, the general solution and maximum radius of a cylindrical construct then become:

$$C(r) = \frac{\varphi}{4D}(r^2 - R^2) + C_o \quad (12)$$

$$R_{max} = \sqrt{\frac{4C_oD}{\varphi}} \quad (13)$$

A comparison of multidimensional steady state solutions with the same parameters is shown in Figure 2.

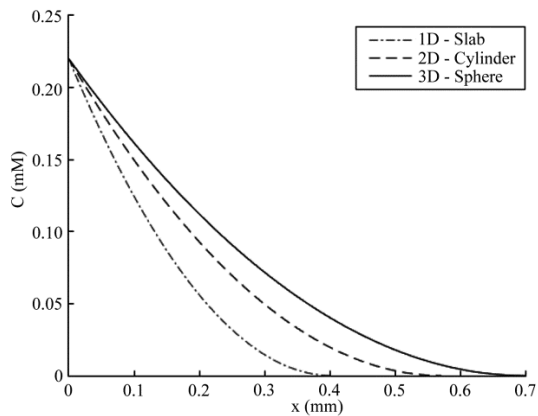

Figure 2: Comparison of metabolic steady-state profiles of oxygen for thickness ($T$) or radius ($R$) in mm in 1D, 2D, and 3D models, each with identical values for $\varphi$, $D$, and $C_o$ (with an inverted spatial axis for the radial cases for comparison).

### 1.3 General Properties of a Tissue Diffusion System

The value of the diffusivity coefficient ($D$) will depend on the characteristics of both the diffusant and the medium through which the molecule diffuses. These characteristics may include nutrient properties like size, charge, and mass[50]; likewise, relevant medium properties may include variations in material density, cellular density, and temperature, all of which can be modeled in non-linear fashion. Nevertheless, it may generally be assumed that the diffusivity coefficient remains a constant parameter in a tissue construct of uniform material density, meaning it is independent of factors such as changing concentration of solutes or changing cellular demand[43,51]. The models presented here also do not include swelling or shrinking of cells, tissues, or hydrogel material or other poroelastic effects, and the assumption of homogenous and isotropic tissue allows derivation of solutions that apply broadly to engineering tissue constructs. For various small molecules in hydrogel polymers, the diffusion coefficient is typically around $10^{-9}$ to $10^{-10}$ m²/s, with small molecules like oxygen typically closer to $10^{-9}$ and slightly larger molecules like glucose closer to $10^{-10}$ m²/s (different units of measure in prior studies are normalized here)[44,51-60]. In hydrogels, diffusivity has been reported to be about 50 to 85% of that found in solution, and diffusivity in cartilage has been reported to be about 40% of that found in solution, with diffusion coefficients for small molecules in a similar range of $2x10^{-10}$ to $8x10^{-10}$ m²/s[43,56,61-62]. Larger proteins like albumin or glycosaminoglycans generally have a smaller diffusion coefficient in the range of $10^{-10}$ to $10^{-11}$ m²/s[62-64].

The initial concentrations ($C_o$) of many types of nutrients are generally provided in the media solution, and culture media may vary considerably in its nutrient content, containing physiologic concentrations of glucose or several times higher concentrations in order to compensate for intermittent media changes and lack of continual perfusion or to study certain conditions and cell types (see Table 1). In cell cultures, standard values of diffused oxygen and nutrient concentrations in gas and liquid phases surrounding the tissue construct may generally be assumed, especially when culture media is well-mixed, or also, in the case of gasses diffusing from outside the media, when the layer of media over the tissue construct is thin or equilibrated for long times. The typical value of oxygen concentration in various depths of solution under standard atmospheric pressure and at 37°C has been reported to be between 0.20 to 0.22 mM[44,57,65-68]. For greater depths of stagnant media over a construct, the concentration of oxygen at the surface of the tissue construct (which here, as with other nutrients at a construct surface, is called $C_o$) can be approximated using Fick's first law of diffusion through overlying liquid media, which, as before, states that flux is equal to the diffusion coefficient multiplied by the change in concentration over the change in distance. Using a thin film assumption, this can be written as:

$$J = \frac{-D}{l}\left(\frac{P}{H} - C_o\right) \quad (14a)$$

where $J$ represents the net flux of oxygen in the $x$-direction, $l$ is the length or thickness of the media over the construct surface (e.g., in length units that match those used in the diffusion coefficient), and the oxygen concentration at the media surface is given by $P/H$ where $P$ is the partial pressure of oxygen (e.g., in units of $atm$) and $H$ is Henry's constant for the media solution (e.g., in units of $\frac{L \cdot atm}{mol}$). Assuming quasi-steady state conditions where $\varphi$ is the metabolic rate of the cellular construct (in units of $\frac{mol}{L \cdot s}$), substituting the flux from Eq. 14a into Eq. 2, and solving for concentration through the liquid media then gives:

$$C = \frac{P}{H} - \frac{\varphi l x}{D} \quad (14b)$$

The value of $C_o$ is found when $x = l$, and $D$ is the diffusion coefficient of the gas in the surrounding liquid media, typically around $3x10^{-9}$ m²/s for oxygen at physiologic temperature and pressure[43-44].

The rate of oxygen consumption ($m_{O_2}$) of neuronal cells in brain across species has been reported to be approximately $1.1x10^{-17}$ to $1.0x10^{-15} \frac{mol}{cell \cdot s}$[69-71] (Table 2). The rate of glucose consumption ($m_{gluc}$) in neuronal cells has been reported to average from about $8x10^{-17}$ to $4x10^{-16} \frac{mol}{cell \cdot s}$ across a variety of rodent and primate species, with a lower range of $1.0x10^{-17}$ to $2.2x10^{-17} \frac{mol}{cell \cdot s}$ for cerebellar neurons and a higher range of $2.0x10^{-16}$ to $4.0x10^{-16} \frac{mol}{cell \cdot s}$ for cortical neurons (Table 3). In humans, cerebellar neurons consume glucose at an average rate of about $1.2x10^{-17} \frac{mol}{cell \cdot s}$ while cortical neurons consume glucose at an average rate of about $2.2x10^{-16} \frac{mol}{cell \cdot s}$[69]. The metabolic rate of whole brain among species varies as a function of brain size and neuronal density, but the metabolic rate per cell is remarkably similar among species, as has been noted previously[69]. These rates can be adapted for the density of cells within a tissue construct by using the formula $\varphi = m\rho$, where $m$ is the metabolic rate for the nutrient for each cell (e.g., in units of $\frac{mol}{cell \cdot s}$) and $\rho$ is the average density of cells in the hydrogel or tissue (e.g., in units of cells/L). This formula is based on the premise that cells are homogeneously distributed or that metabolic consumption values are similar throughout the construct, a premise which is generally valid for relatively uniform structures like neurospheres and embryoid bodies[72]. Thus, as an example, for a construct with 20,000 neuronal cells in 25 µl of hydrogel, $\rho = 8.0x10^8$ cells/L, and the final value of $\varphi$ for the construct becomes about $9.8x10^{-9}$ to $8.8x10^{-7} \frac{mol\,O_2}{L \cdot s}$ for oxygen. Of note, the metabolic consumption by the entire construct can also be found by multiplying $\varphi$ by the volume of the 3D tissue construct (e.g., $V_c = \frac{4}{3}\pi r^3$ for a sphere), which can be useful in determining the perfusion needed or the amount and frequency of media changes for nutrients supplied by the media.

The minimal and maximal thickness of hydrogel can then be estimated via these parameters and formulas. If minimizing metabolic parameter values are used—that is, if neuronal cells described above consume oxygen at an upper range of $8.8x10^{-7} \frac{mol\,O_2}{L \cdot s}$ and if the diffusion coefficient is at the lower range of $10^{-10}$ m²/s—then the viable thickness of tissue is predicted to be only about 250 μm. Conversely, if the maximizing parameters are used, meaning oxygen consumption of $9.8x10^{-9} \frac{mol\,O_2}{L \cdot s}$ and diffusion coefficient of $10^{-9}$ m²/s, then the maximum thickness of viable tissue under these idealized culture conditions could be as much as 16 mm assuming steady state with unconstrained flux at the construct surface. For the average human neuron metabolic rate of oxygen and at an intermediate diffusivity, the viable slab thickness is predicted to be 1.4 mm, with a range of 0.4 to 2 mm for minimal and maximal diffusivities respectively, or thicker for more quiescent cells, meaning that the viability of neurons in 3D tissue depends significantly on the neural activity and the proximity and diffusivity of nutrients in tissue. This thickness limit for cell viability has been observed experimentally in layered hydrogel constructs with differentiated human neurons with single surface layer diffusion that were capable of being up to 2 mm thick[32]. Similarly, analysis of the spherical equation for a construct with 20,000 neurons in 25 μl of hydrogel provides corroborating results, with the viable diameter being 0.9 to 4.1 mm for minimal and maximal diffusivities respectively. This means that a neuronal sphere with normal spontaneous neural activity and moderate metabolic activity and a diameter of about 4 mm should just be viable at this cell density, and it is interesting to note that this was the size limit reported in cerebral organoids made from pluripotent stem cells differentiated within spherical hydrogels, although the cellular density in these organoids was unknown[31].

## 2.1 Diffusion Models (Steady and Non-Steady-States) in the Infinite Domain

The diffusion equation can be solved through several well-known methods if it is assumed that diffusion occurs through an infinite spatial distance. Because these solutions are only applicable over an infinite spatial range, however, they are not particularly helpful for modeling diffusion and metabolism in finite constructs, but they are helpful for understanding the general concepts of diffusion solutions. Solution methods include the use of LaPlace transforms or, through the strategy of Boltzmann's transformation, substituting $\mu \overset{\text{def}}{=} \frac{x}{\sqrt{4Dt}}$ into Eq. 1 to make the partial differential equation into an ordinary differential equation[42], and solving through a series of steps with desired initial conditions.

$$-2\mu \frac{\partial C}{\partial \mu} = \frac{\partial^2 C}{\partial \mu^2} \quad (15)$$

When integrated twice with the initial condition $C(x, t = 0) = C_o$ over the semi-infinite distance ($0 < x < \infty$), the result is[42,52,73]:

$$C(\mu) = C_o[1 - \text{erf}(\mu)] \quad (16)$$

where

$$\text{erf}(\mu) = \frac{2}{\sqrt{\pi}} \int_0^\mu e^{-\vartheta^2} d\vartheta \quad (17)$$

wherein the variable $\vartheta$ acts as a place holder for $\mu$ over the integration. It can be verified that this is indeed a solution by substituting back into the original Eq. 1. Finally, substituting back for $\mu$ gives the semi-infinite solution in terms of $x$:

$$C(x, t) = C_o \left[1 - \text{erf}\left(\frac{x}{\sqrt{4Dt}}\right)\right] \quad (18)$$

Although the error function (erf) appears complex, it can be solved by many types of programs, including by simply typing "erf(  )" into the website google.com. This provides a solution in the infinite spatial domain for diffusants that are not consumed by the cells and that are held constant in the surrounding solution (the condition $(0, t) = C_o$ ). It can be seen here that the value $\mu = \frac{x}{\sqrt{4Dt}}$ is a dimensionless parameter, which allows comparison of diffusion behavior across different times and diffusivities. The time required for the concentration to reach a given value at a given point is proportional to the square of the distance into the construct and also inversely proportional to the diffusivity. Values for desired time (t) and diffusivity (D) can be inserted and the concentration profile solved and graphed (Figure 3e-f). Under the above assumption that the amount of oxygen and nutrient outside the hydrogel remains constant, this model shows that the concentration of diffusant reaches 50% of $C_o$ through 1 mm of hydrogel in about 20 minutes and 90% of $C_o$ in approximately 8.8 hours when D=10⁻⁹ m²/s. This is in accordance with other reports of an 8 hour diffusion time for small molecules over a distance of 1 mm in hydrogel[74].

The diffusion equation can also be solved over the semi-infinite distance for diffusant substances that are not consumed by the cells and that only have a finite supply in the media (like certain signaling molecules or growth factors). At $t=0$, the concentration of diffusant from media into hydrogel appears as a step function at x=0, where concentration in the media is $C_o$ and concentration in the hydrogel is zero. The solution can be found by treating the finite amount of diffusant at each position as a source of diffusion to adjacent positions, and letting $N$ represent the total number of diffusant particles in the source solution, which in the one-dimensional case is[42]:

$$N = \int C_o \, d\xi \quad (19)$$

In this system, the variable $x$ is treated as the position where concentration is measured and $\xi$ is the position acting as source of diffusant. The solution over the semi-infinite domain is then the integration over all sources along $\xi$,

$$C(x,t) = \frac{N}{\sqrt{\pi Dt}} \int_0^\infty e^{\left[\frac{-(\xi-x)^2}{4Dt}\right]} \quad (20)$$

producing a solution for $C$ in terms of $x$ and $t$ for the finite nutrient over semi-infinite distance[42] (results shown in Figure 3(a-d)):

$$C(x,t) = \frac{N}{\sqrt{\pi Dt}} e^{-\frac{x^2}{4Dt}} \quad (21)$$

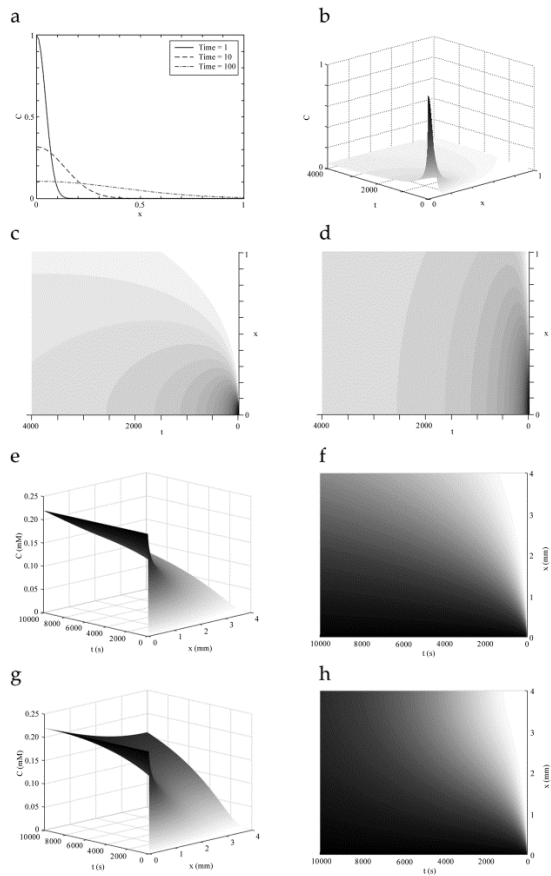

Figure 3: Concentration profiles in a semi-infinite domain of constant diffusivity as a function of distance $x$ (depth through hydrogel in mm) and time $t$ (in seconds). Having assumed a finite amount of diffusant (like glucose in a culture well, where D=10⁻¹⁰ m²/s, with $C_o$ set at unity) in an infinite domain, part (a) shows the general case of how concentration changes as a function of distance in the semi-infinite domain, with time shown as sequential curves. Part (b) shows the general case of concentration of a finite nutrient diffusing in space and time, as described by Eq. 21, and as with all 3D graphs herein, depth is shown on the rightward axis and time on the leftward axis. The concentration intensity profile from part (b) is shown as a contour plot in distance and time for either (c) D=10⁻¹⁰ m²/s or (d) D=10⁻⁹ m²/s, all other parameters being equal. Time is shown until 4,000 s in parts (b-d). Parts (e-f) show the condition where, rather than finite diffusant, external $C_o$ remains constant (like ambient oxygen), as described in the semi-infinite domain of Eq. 18. Parts (g-h) demonstrate diffusion and concentration profile in a 4 mm thick finite construct with unlimited supply of oxygen as described in Eq. 22. In parts (e-h), D=10⁻⁹ m²/s, $C_o$=0.22 mM, and time is shown until 10,000 s.

*3.1 Diffusion Models of Finite Slab Constructs*

Interestingly, similar transformations and solutions using the error function can be used to represent diffusion into a tissue construct of any thickness in the finite domain. Solutions for general diffusion along the *x*-axis into a slab have been derived for various applications[41,45], which may be modified here to apply to the particular case of tissue constructs of thickness $T$:

$$C(x,t) = C_o \left[ \text{erfc}\left(\frac{x}{\sqrt{4Dt}}\right) - \sum_{n=1}^{\infty} (-1)^{n+1} \text{erfc}\left(\frac{(2n)T - x}{\sqrt{4Dt}}\right) + \sum_{n=1}^{\infty} (-1)^{n+1} \text{erfc}\left(\frac{(2n)T + x}{\sqrt{4Dt}}\right) \right] \quad (22)$$

In this model, concentration continues to rise even at the deepest point within the construct as long as the nutrient supply remains higher at the source point, as can be seen in the case of oxygen diffusion into a 4 mm thick slab as graphed in Figure 3(g-h). However, this solution still does not completely describe diffusion limitations or consumption of nutrients by cells within the tissue construct.

*3.2 Steady-State Diffusion Profiles in Metabolically-Maximized Tissue Constructs*

By substituting the formulas for maximal diffusion depth (Eq. 4, 10, or 13) into the metabolic concentration profiles for each form of construct (Eq. 3, 9, or 12), general steady-state solutions for metabolically-maximized constructs can be obtained and written in simplified form:

$$\text{Slab:} \quad C(x) = \frac{\varphi x^2}{2D} - \sqrt{\frac{2C_o \varphi}{D}} x + C_o \quad (23a)$$

$$\text{Cylinder:} \quad C(r) = \frac{\varphi r^2}{4D} \quad (23b)$$

$$\text{Sphere:} \quad C(r) = \frac{\varphi r^2}{6D} \quad (23c)$$

These formulas describe the steady-state metabolic profile through the construct, and it may be verified that these solutions meet the assigned conditions of $C = C_o$ at $x = 0$ and $C = 0$ at $x = T_{max}$ for the slab, and $C = 0$ at $r = 0$ and $C = C_o$ at $r = R_{max}$ for the two radial cases. These solutions, however, do not describe the transient phase of unsteady state where concentration profile through the construct progresses towards the steady state from some initial value. The initial concentration of oxygen and nutrients in a newly formed tissue construct is typically zero ($C_i = 0$), such as when a hydrogel is mixed and cellularized outside of media or under anoxic conditions before cellularization, as is often done with many types of polymer hydrogels, or when scaffolds are seeded and transferred to new media, or when a new nutrient is introduced during stages of induction and differentiation. It has been noted that a large amount of cell death occurs in the early initial period immediately after the seeding of a new 3D tissue construct[75], which is likely due to these nutrient deficiencies, and which suggests that an understanding of how concentrations change within a tissue construct during transient perturbations is especially important in tissue engineering.

*3.3 Steady and Non-Steady State Model of Slab Constructs Constrained to Maximization with Unlimited Nutrient*

In order to derive non-steady-state solutions for constructs with diffusion and metabolism, it is instructive to first derive solutions for the simplest condition for a theoretical maximized construct, where the surface or starting point concentration is held constant (e.g., $C = C_o$) and where the maximal depth concentration is held constant (e.g., $C = 0$), as would be expected in a maximized construct where all nutrient was completely consumed by the deepest point at steady-state. Boundary conditions with constant values are generally known as Dirichlet boundary conditions, and this form of solution is convenient both because it provides a foundation for how more complicated solutions will be derived and also because there are many scenarios where the value of each parameter in a construct is not known, meaning that maximal theoretical depth is not known, but because a maximal depth for a viable tissue construct can be observed empirically, this value can simply be inserted in lieu of the individual parameters that affect such limits. Thus this unique approach can still allow an estimation of diffusion dynamics within the tissue construct even when parameters of diffusivity, initial concentrations of diffusant, cellular density, or metabolic rates of cells are not known. Maximized tissue constructs are also generally of the most interest in research applications since they represent the limits of tissue size, metabolic activity, and cell viability, and this method can also allow individual parameters to be interpolated from experimental data or allow modeling of maximal diffusion depth that changes over time or areas of nutrient surplus and deficiency. Such forms of solutions are also useful where consumption of nutrient is maintained by external factors other than cell consumption through the construct, such as when a nutrient source is supplied at one side of a tissue construct and washed out at an opposing side. Such a scenario may occur, for example, when cells are fed with one nutrient media over the tissue surface but then perfused with a separate media lacking the nutrient underneath the cells in order to create a gradient through the tissue, or when the nutrient undergoes degradation, sequestration, chelation, absorption, or reaction when exposed at one side of the tissue construct. Solutions where the nutrient molecule is consumed by cells throughout the construct will then be presented in more detail in subsequent sections.

Complete solutions over a finite domain may be obtained by means of a method of separation of variables with the application of specific boundary conditions, a method that was developed for this form of equation by Joseph Fourier in the early 19th century[76]. This method assumes that the solution for $C(x,t)$ equals the product of a function of $x$ and a function of $t$, defined as F($x$) and G($t$) respectively.

$$C(x,t) = F(x)G(t) \quad (24)$$

The two functions F($x$) and G($t$) can be plugged into the diffusion equation (1) to give:

$$\frac{1}{DG}\frac{dG}{dt} = \frac{1}{F}\frac{d^2F}{dx^2} \quad (25)$$

Because the two different functions on the left and right sides of the equation are equal, a solution can be found by setting each side of the equation equal to an arbitrary constant $\gamma$ and creating a set of ordinary differential equations (ODEs) that can be solved with the appropriate boundary conditions over the finite domain:

$$\begin{cases} \dfrac{dG}{dt} = \gamma DG \\ \dfrac{d^2F}{dx^2} = \gamma F \end{cases} \quad (26)$$

The critical condition of cells consuming exactly enough nutrient that concentration approaches zero at the deepest point in the construct is of particular interest since it describes a maximal tissue size and allows Dirichlet boundary conditions to conveniently be set at both the surface and depth of the tissue construct, as in $C(T,t) = 0$. The case of a constant supply of oxygen or nutrient at the construct interface is first considered here, meaning $C(0,t) = C_o$. As mentioned previously, the concentration of oxygen and nutrients in the construct at initiation is often zero due to the fact that many tissue constructs are created by seeding cells into hydrogels or scaffolds, meaning $C(x,0) = 0$ for the range $0 \le x \le T$ at $t = 0$. If $\gamma > 0$ (which can be defined by $\gamma = \varepsilon^2$, where $\varepsilon$ is the range of eigenvalues $\frac{n\pi}{T}$, sometimes called Helmholtz eigenvalues after the physicist and physician Hermann von Helmholtz who investigated the form of equation for $F$ above), then $F(x) = ae^{\varepsilon x} + be^{-\varepsilon x}$, which under the boundary conditions above simply results in the uninteresting case of $a = b = 0$. If $\gamma = 0$, then $F(x) = ax + b$, which under the boundary conditions results in $F(0) = C_o = b$ and $a = -\frac{C_o}{T}$, meaning $F(x) = C_o(1 - \frac{x}{T})$, and $\frac{dG}{dt} = 0$. This provides a steady state solution, but because $G(t)$ is a constant in this case, this solution alone does not adequately describe the change in concentration over time. The non-steady-state solution is found when $\gamma < 0$, and the general form of the solutions become $F(x) = a\cos\left(\frac{n\pi}{T}x\right) + b\sin\left(\frac{n\pi}{T}x\right)$ and $G(t) = ge^{-\left(\frac{n\pi}{T}\right)^2 Dt}$ where $a$, $b$, and $g$ are found from the conditions of the system: $F(0) = 0 = a$, $F(T) = b\sin(\varepsilon T)$, meaning that $F(x) = b\sin(\frac{n\pi}{T}x)$. It may be noted that a high Biot coefficient ($B$) describes a system where the mass transfer coefficient ($h$) at the surface of the construct is much larger than the diffusivity ($D$), meaning that a molecule is inhibited at a phasic interface only by diffusivities of the medium and there is not a prescribed time lag for a molecule crossing at a phasic interface, as described by the formula $B = \frac{hR}{D}$, where $h$ and $D$ are given in the condition $-D\frac{dC}{dr} = hC$ at the construct surface $r = R$, and also indicating that the system is defined by Dirichlet-type boundary conditions rather than Neumann-type. By the principle of superposition then, the linear combination of solutions of $F(x)$ and $G(t)$ can then be given by Fourier series representation of the eigenfunctions:

$$C(x,t) = C_o\left(1 - \frac{x}{T}\right) + \sum_{n=1}^{\infty} ge^{-\left(\frac{n\pi}{T}\right)^2 Dt}\sin\left(\frac{n\pi}{T}x\right) \quad (27)$$

By applying the boundary condition $C(x,0) = 0$, again meaning that at $t = 0$ no diffusion has yet occurred, then the value of the odd function $g$ can be found by using the orthonormality property of eigenfunctions and multiplying $C(x)$ by $F(x)$ and integrating over the range of $x$, or by using the related solution given in standard Fourier series tables[77]:

$$g = \frac{2}{T}\int_0^T -C_o\left(1-\frac{x}{T}\right)\sin\left(\frac{n\pi}{T}x\right)dx = \frac{-2C_o}{n\pi} \quad (28)$$

It may be noted that the Biot coefficient ($B$) can be described by the formula $B = \frac{hT}{D}$, where $h$ and $D$ are given in the condition $-D\frac{dC}{dx} = hC$ at the construct surface $x = T$, and a high Biot coefficient describes a system where the mass transfer coefficient ($h$) at the surface of the construct is much larger than the diffusivity ($D$), meaning that a molecule is inhibited at a phasic interface only by diffusivity of the medium and there is not a prescribed time lag for crossing at the interface, and also indicating that the system is defined by Dirichlet-type boundary conditions rather than Neumann-type. Applying the boundary conditions of 1) $C(x=0,t) = C_o$, 2) $C(x=T,t) = 0$, & 3) $C(x,t=0) = 0$, a final complete solution for $C(x,t)$ in a maximized finite construct of thickness $T$ thus becomes:

$$C(x,t) = C_o\left(1-\frac{x}{T}\right) - \frac{2C_o}{\pi}\sum_{n=1}^{\infty}\frac{1}{n}e^{-\left(\frac{n\pi}{T}\right)^2 Dt}\sin\left(\frac{n\pi}{T}x\right) \quad (29a)$$

Through a method of LaPlace transforms, this solution can also be conveniently written in the form of error functions:

$$C(x,t) = C_o\left[\text{erfc}\left(\frac{x}{\sqrt{4Dt}}\right) - \sum_{n=1}^{\infty}\text{erfc}\left(\frac{(2n)T-x}{\sqrt{4Dt}}\right) + \sum_{n=1}^{\infty}\text{erfc}\left(\frac{(2n)T+x}{\sqrt{4Dt}}\right)\right] \quad (29b)$$

The maximal tissue construct thickness obtained from Eq. 4 can be inserted for $T$, which produces a complete closed-form solution for concentration within a construct in space and time. This solution was graphed in MATLAB as shown in Figure 4(a-b), and comparison to numerical PDE solutions in MATLAB demonstrated exact matches for all equivalent conditions (see Appendix for PDE solver methods). The series can be approximated well for early times with $n$ values of about 500 or less, and later times can be adequately represented with an $n$ of only 10 or less. For conditions where the initial concentration in the construct ($C_i$) is a value other than zero, ($C_o - C_i$) may be substituted for $C_o$ values and ($C - C_i$) substituted for $C$ as appropriate. For a model where $C_i = 0$ and maximal thickness is calculated to be 4 mm (meaning that all remaining oxygen is consumed at the deepest point in the construct at steady state), the concentration of oxygen at the halfway point through the construct reaches 50% of its steady-state value in approximately 25 minutes and 90% of its steady-state value in approximately 1.2 hours.

*3.4 Steady and Non-Steady State Model of Slab Constructs Constrained to Maximization with Limited Nutrient*

For limited nutrient that is consumed by cells from the media (e.g., glucose), it is also important to account for the drop in concentration that occurs in the surrounding media as nutrient is consumed, which in turn feeds back onto the driving source of diffusion. Accurately modeling this behavior with closed-form solutions can be difficult, but the drop in concentration from the media is simply a function of two components: 1) the amount of nutrient consumed by the construct, and 2) the amount of nutrient diffused into the construct. By using the total consumption from the tissue construct and adjusting for the difference in volume between the construct and the media volume, $C_o$ can be replaced with a function that describes the rate of metabolic consumption from the media by the construct, where $V_c$ represents the volume of the tissue construct and $V_m$ is the media volume around the tissue construct:

$$C_{media} = C_o - \varphi t\frac{V_c}{V_m} \quad (30)$$

Secondly, the drop in concentration from the media due to diffusion alone into the construct at any point in time should also be accounted for, represented by $\bar{C}$, with formulas for different construct types given in Eq. 49a-c (although, if desired, for practical simplification this term may also be neglected). Neglecting metabolism for a moment, the molar amount of nutrient in a closed diffusion system remains constant, meaning that:

$$C_o V_m = C_{media} V_m + \bar{C} V_c \quad (31)$$

Solving for $C_{media}$, the drop in concentration in media due to diffusion alone is described by:

$$C_{media} = C_o - \bar{C} \frac{V_c}{V_m} \quad (32)$$

The formulas for $\bar{C}$ in Eq. 49a-c, however, are for a constant unlimited nutrient, and when a limited nutrient diffuses from media into a construct, $\bar{C}$ cannot rise all the way to $C_o$ since the volume $V_m + V_c$ is greater than $V_m$ alone. Rather, $\bar{C}$ reaches a new equilibrium in both media and construct, which can be expressed by replacing $C_o$ in Eq. 49a-c with $\frac{C_o V_m}{V_m + V_c}$, and insertion of this expression along with Eq. 49a-c for $\bar{C}$ in Eq. 32 produces a formula for the drop in concentration from media due to diffusion that accounts for the total volume of $V_m + V_c$:

$$C_{media} = C_o - \bar{C} \frac{V_c}{V_m + V_c} \quad (33)$$

Finally, a complete solution for decline of limited nutrient from media due to diffusion and metabolic consumption may then be formed by combining the metabolic consumption and the diffusional compartment shift decrease, replacing $C_o$ in Eq. 29 with:

$$C_{media} = \left( C_o - \varphi t \frac{V_c}{V_m} - \bar{C} \frac{V_c}{(V_m + V_c)} \right) \quad (34)$$

It is important to note that in the case of limited nutrient where $C_o$ is consumed in time, the maximized value of $T$ given by Eq. 4 also becomes a function of time,

$$T_{max_t} = \sqrt{\frac{2sD \left( C_o - \varphi t \frac{V_c}{V_m} - \bar{C} \frac{V_c}{(V_m + V_c)} \right)}{\varphi}} \quad (35)$$

which altogether produces the following model for concentration in maximized constructs with limited nutrient:

$$C(x,t) =$$
$$\left( C_o - \varphi t \frac{V_c}{V_m} - \bar{C} \frac{V_c}{(V_m + V_c)} \right) \left( 1 - \frac{x}{\sqrt{\frac{2D\left(C_o - \varphi t \frac{V_c}{V_m} - \bar{C} \frac{V_c}{(V_m + V_c)}\right)}{\varphi}}} \right) - \frac{2\left(C_o - \varphi t \frac{V_c}{V_m} - \bar{C} \frac{V_c}{(V_m + V_c)}\right)}{\pi} \sum_{n=1}^{\infty} \frac{1}{n} e^{\frac{-(n\pi)^2 \varphi t}{2\left(C_o - \varphi t \frac{V_c}{V_m} - \bar{C} \frac{V_c}{(V_m + V_c)}\right)}} \sin \left( \frac{n\pi x}{\sqrt{\frac{2D\left(C_o - \varphi t \frac{V_c}{V_m} - \bar{C} \frac{V_c}{(V_m + V_c)}\right)}{\varphi}}} \right) \quad (36)$$

This solution is valid within the domain of $T_{max_t}$ and until $C_o$ reaches zero in media (given by Eq. 51). The solution shows that the concentration will have a maximal peak at a slightly different time at each point in the construct. Using $C_o$= 10 mM, D=10⁻¹⁰ m²/s, and $\varphi$ for glucose consumption of $1.25x10^{-7} \frac{mol}{L \cdot s}$, the local maximum at 50% distance into the 4 mm maximal thickness of the construct occurs at 11.6 hours with a peak concentration 35% of initial $C_o$, while at 90% depth into the construct the local maximum occurs at 8.5 hours with a peak concentration 4% of initial $C_o$. Assuming media volume is only five times that of the construct volume, glucose in the media is completely consumed by 93 h (Figure 4(c-d)). At early times the zero order consumption plus diffusional equilibration causes an exponential-like decay in nutrient concentration at the construct surface, while at later times once the majority of diffusion has occurred the decline becomes linear as would be expected with zero order consumption. The maximal viable thickness of the construct changes as time progresses since the outer concentration of nutrient is changing with time, meaning that nutrient deficiency will occur both in transient states and as nutrient is consumed. The $T_{max}$ can also be modified to account for changes in $\varphi$, such as when cells proliferate and change cell density in the tissue construct.

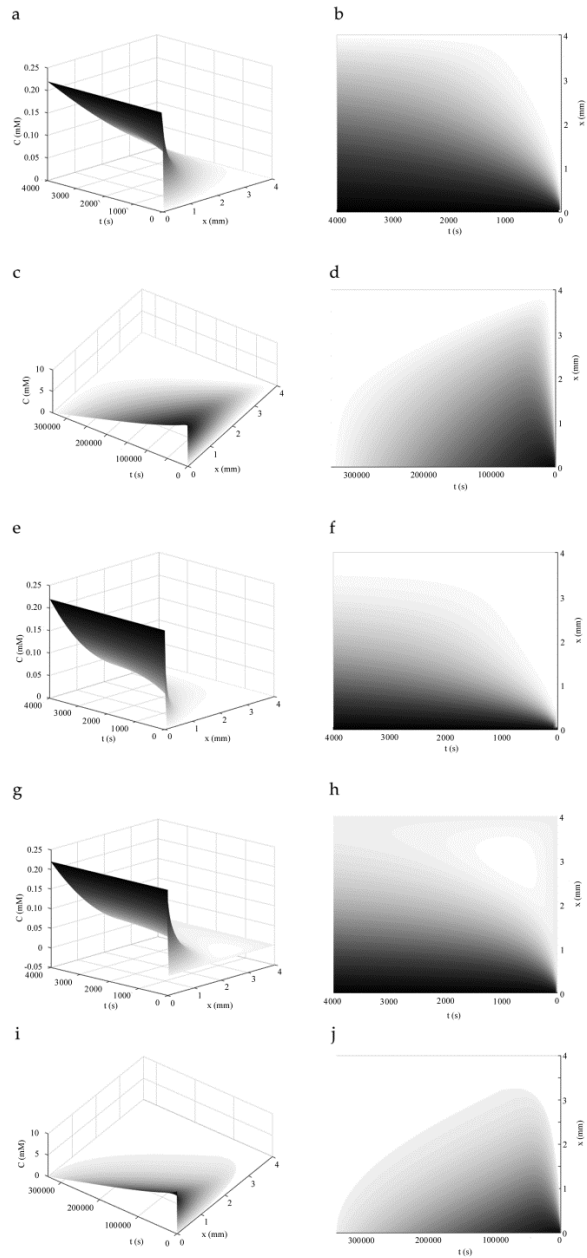

Figure 4: (a-b) Concentration profiles for oxygen in culture media using the constrained solution for unlimited nutrient (Eq. 29) over a construct's spatial domain: $C_o$=0.22 mM, D=10$^{-9}$ m²/s, $T_{max}$=4 mm, and time shown from 0 to 4000 seconds. The view of all axes is shown in (a) and the concentration intensity profile in time and distance (horizontal and vertical axes respectively) is shown in part (b). In (c-d) the concentration profile of glucose from the solution in Eq. 36 is shown using $C_o$=10 mM, D=10$^{-10}$ m²/s, and $\varphi = 1.25x10^{-7}\ \frac{mol}{L \cdot s}$ for $T_{max}$=4 mm, until total consumption at t=3.33x10$^5$ s at x=0 with a media volume five times that of the construct volume. Under these conditions the ramp up in concentration due to diffusion occurs much more quickly than the consumption of nutrient. In part (e), the transient ramp up in oxygen concentration within the construct is shown for the first 4000 s as it approaches parabolic quasi-steady state, as described by the solution in Eq. 37 using $C_o$=0.22 mM, D=10$^{-9}$ m²/s, $\varphi_{O_2} = 2.75x10^{-8}\ \frac{mol}{L \cdot s}$, and $T_{max}$=4 mm. The concentration profile of oxygen in the construct over time is shown in (f). The numerical solution with the same conditions as (e-f) is shown in (g-h), as explained in the text. The concentration profile of glucose in the construct from the metabolic solution in Eq. 38 is shown in (i-j) over the valid domain of $x = 0$ to $T_{max}$ and until total consumption at 3.3x10$^5$ s using $C_o$=10 mM, D=10$^{-10}$ m²/s, $\varphi = 1.25x10^{-7}\ \frac{mol}{L \cdot s}$, and $T_{max}$=4 mm, with a media volume five times that of the construct volume.

*3.5 Steady and Non-Steady State Model of Metabolically-Maximized Slab Constructs with Unlimited Nutrient*

The prior solutions, however, do not account for the effects of metabolism by cells homogenously distributed through a construct. For any case where metabolism of nutrient by cells is greater than zero ($\varphi > 0$), the concentration profile for the nutrient through the construct will be parabolic, as described previously. A complete steady and non-steady state model for diffusion and metabolism with unlimited nutrient may be derived as before by again implementing superposition of particular solutions that satisfy the boundary conditions of 1) $C(x=0,t) = C_o$, 2) $C(x = T_{max}, t) = 0$, and the initial state of 3) $C(x, t=0) = 0$, and that satisfy the known steady-state solution (Eq. 3):

$$C(x,t) = C_o + \frac{\varphi x^2}{2D} - \frac{\varphi T x}{D} - \frac{2C_o}{\pi}\sum_{n=1}^{\infty}\frac{1}{n}e^{-\left(\frac{n\pi}{T}\right)^2 Dt}\sin\left(n\pi\varphi x\frac{2T-x}{2C_o D}\right) \quad (37)$$

The $T_{max}$ value, either calculated from parameters or observed empirically, can then be inserted for $T$. A graph of Eq. 37 with typical oxygen parameters is shown in Figure 4(e-f). Comparison of Eq. 37 to numerical solutions under the same conditions shows that the analytic and numerical solutions are identical at steady-state, but there is a small variance at early transient times in the construct, where concentration briefly becomes negative in numerical models due to the construct having a constant $\varphi$ but nutrient not having yet reached the full depth of the construct, as demonstrated in Figure 4(g-h). This transient negative deficiency is only present in the numerical models due to the assumed condition that consumption of nutrient continues even in the absence of nutrient, thereby building up a deficiency until nutrient arrives, as opposed to the analytic models which keep nutrient concentration at the initial zero value until nutrient has diffused to the space, as would be measured in a culture system. Yet the analytic models presented herein still closely replicate the numerical models with all appropriate boundary conditions and may in fact represent more realistic conditions where cells cannot metabolize nutrient until the nutrient arrives, where nutrient concentration cannot be measured as a negative value in culture, and where cells do not simply restore nutrient deprivation as a linear summation over time. Mathematical code for both the numerical and analytical approaches is provided for convenience in the Appendix.

### 3.6 Steady and Non-Steady State Model of Metabolically-Maximized Slab Constructs with Limited Nutrient

As before, for limited nutrient that is consumed by cells from the media, $C_o$ can be replaced with Eq. 34, where $\bar{C}$ is given by Eq. 49a and the maximized value of $T$ becomes a function of time (Eq. 35), forming a complete model for concentration within the construct of limited nutrient that is consumed from the media:

$$
\begin{aligned}
C(x,t) = \\
\left(C_o - \varphi t \frac{V_c}{V_m} - \bar{C}\frac{V_c}{(V_m + V_c)}\right) + \frac{\varphi x^2}{2D} - x\sqrt{\frac{2\varphi\left(C_o - \varphi t \frac{V_c}{V_m} - \bar{C}\frac{V_c}{(V_m + V_c)}\right)}{D}} \\
- \frac{2\left(C_o - \varphi t \frac{V_c}{V_m} - \bar{C}\frac{V_c}{(V_m + V_c)}\right)}{\pi}\sum_{n=1}^{\infty}\frac{1}{n}e^{-\left(\frac{n\pi}{T}\right)^2 Dt}\sin\left(n\pi x\frac{\sqrt{8D\varphi\left(C_o - \varphi t \frac{V_c}{V_m} - \bar{C}\frac{V_c}{(V_m + V_c)}\right)} - \varphi x}{2D\left(C_o - \varphi t \frac{V_c}{V_m} - \bar{C}\frac{V_c}{(V_m + V_c)}\right)}\right) \quad (38)
\end{aligned}
$$

The result of Eq. 38 with typical glucose parameters is shown in Figure 4(i-j).

### 4.1 Diffusion Models of Finite Cylindrical Constructs

By the same method of separation of variables used to derive the one-dimensional solution, the general equation can be obtained for the cylindrical case, where again $F$ is a function of radial distance and $G$ is a function of time:

$$\frac{1}{DG}\frac{dG}{dt} = \frac{1}{rF}\frac{d}{dr}\left(r\frac{dF}{dr}\right) \quad (39)$$

The general solutions to the form of $F(r)$ in this case involve Bessel functions, and with the boundary conditions of constant concentration of nutrient in the media, $C(R,t) = C_o$, and initial concentration of zero within the cylinder, $C(r,0) = 0$ for the range $0 \le x \le R$, as well as the assumption that the length of the cylinder is significantly greater than the radius of the cylinder, the solution is found by previously described methods[39,41,45]

$$C(r,t) = C_o\left[1 - \frac{2}{R}\sum_{n=1}^{\infty}e^{-(\lambda_n)^2 Dt}\frac{J_o(r\lambda_n)}{\lambda_n J_1(R\lambda_n)}\right] \quad (40)$$

where $J_o$ and $J_1$ are Bessel functions of the first kind (orders zero and one respectively), and where the eigenvalues $\lambda_n$ are the roots of $J_o(R\lambda_n) = 0$. By using the values in Table 4, where the first ten roots are shown, $\lambda_n$ can be solved and inserted in the summation series. Bessel functions can be approximated with a simplified formula[78]:

$$J_0(x) \approx \sqrt{\frac{2}{\pi x}} \left(1 - \frac{1}{16x^2} + \frac{53}{512x^4}\right) \cos\left(x - \frac{\pi}{4} - \frac{1}{8x} + \frac{25}{384x^3}\right) \quad (41a)$$

$$J_1(x) \approx \sqrt{\frac{2}{\pi x}} \left(1 + \frac{3}{16x^2} - \frac{99}{512x^4}\right) \cos\left(x - \frac{3\pi}{4} - \frac{3}{8x} - \frac{21}{128x^3}\right) \quad (41b)$$

Therefore, although quite unwieldy, an approximate closed-form solution for cylindrical diffusion can be written as follows, where values for $\lambda_n$ are given in Table 4:

$$C(r,t) = C_o \left[ 1 - \frac{2}{R} \sum_{n=1}^{\infty} e^{-(\lambda_n)^2 Dt} \frac{\sqrt{\frac{2}{\pi r \lambda_n}} \left(1 - \frac{1}{16(r\lambda_n)^2} + \frac{53}{512(r\lambda_n)^4}\right) \cos\left(r\lambda_n - \frac{\pi}{4} - \frac{1}{8r\lambda_n} + \frac{25}{384(r\lambda_n)^3}\right)}{\sqrt{\frac{2\lambda_n}{\pi R}} \left(1 + \frac{3}{16(R\lambda_n)^2} - \frac{99}{512(R\lambda_n)^4}\right) \cos\left(R\lambda_n - \frac{3\pi}{4} - \frac{3}{8R\lambda_n} - \frac{21}{128(R\lambda_n)^3}\right)} \right] \quad (42)$$

The primary weakness in this approach, however, is that the approximations of Bessel functions become inaccurate at small values of $r$, and the sum function is inaccurate for small values of $t$ unless carried out to large $n$ values, meaning that at early times in the center of the sphere this simplified closed-form solution falls apart. However, by using actual Bessel function calculations in MATLAB, more accurate models for general cylindrical diffusion can be achieved (Figure 5(a-d)).

### 4.2 Steady and Non-Steady State Model of Metabolically-Maximized Cylindrical Constructs with Unlimited Nutrient

As with the 1D case, the 2D cylindrical case also uses the steady state defined by metabolism in the construct (Eq. 12). The cylindrical equation for a maximized value of $R$ that results in zero concentration at the center of the cylinder is implemented with boundary conditions of 1) $C(r = R_{max}, t) = C_o$, 2) $C(r = 0, t) = 0$, and the initial state of 3) $C(r, t = 0) = 0$, and a complete steady and non-steady state model becomes:

$$C(r,t) = \frac{C_o r^2}{R^2} + \frac{2C_o}{\pi} \left[ \sum_{n=1}^{\infty} \frac{-1^n}{n} e^{-\left(\frac{n\pi}{R}\right)^2 Dt} \sin\left(n\pi \frac{r^2}{R^2}\right) \right] \quad (43)$$

Graphical representation of this equation is shown in Figure 5(e-f). It is useful to note that in the radial cases, $\frac{C_o r^2}{R_{max}^2}$ equals $\frac{\varphi r^2}{4D}$ for the cylinder (Eq. 23b) or $\frac{\varphi r^2}{6D}$ for the sphere (Eq. 23c). If desired, the concentration profile through the construct may be compared to the 1D slab case by reversing the spatial axis via substitution of $(R_{max} - r)$ in place of $r$. The numerical solution under the same conditions of Eq. 47 and with the same parameters used in Figure 5(e-f) is shown in Figure 5(g-h), demonstrating a transient negative nutrient concentration at an early phase where cells are presumed to continue nutrient consumption even as nutrient diffuses towards the inner region of the construct, as described in Section 3.5.

### 4.3 Steady and Non-Steady State Model of Metabolically-Maximized Cylindrical Constructs with Limited Nutrient

A model for limited nutrient consumed in the maximized construct can then be created as before by substituting $C_o = \left(C_o - \varphi t \frac{V_c}{V_m} - \bar{C} \frac{V_c}{(V_m + V_c)}\right)$ from Eq. 34 and $R_{max_t} = \sqrt{\frac{4D\left(C_o - \varphi t \frac{V_c}{V_m} - \bar{C}\frac{V_c}{(V_m+V_c)}\right)}{\varphi}}$ from Eq. 35 into Eq. 43 (where $\bar{C}$ is given by Eq. 49b and $R_{max_t}$ is substituted in place of $R$), which simplifies substantially after canceling terms. Because these models describe tissue constructs at a constant thickness or constant radius, and because the maximal depth that changes over time should be measured from the surface of the constructs (as was shown in the 1D case of a slab with limited nutrient), the variable $r$ in radial cases of limited nutrient should also be replaced with $(r - R_{max} + R_{max_t})$, as shown in Eq. 44.

$$C(r,t) =$$
$$\frac{\varphi(r - R_{max} + R_{max_t})^2}{4D} + \frac{2\left(C_o - \varphi t \frac{V_c}{V_m} - \bar{C}\frac{V_c}{(V_m + V_c)}\right)}{\pi} \left[ \sum_{n=1}^{\infty} \frac{-1^n}{n} e^{\left(\frac{-\varphi t (n\pi)^2}{4C_o - 4\varphi t \frac{V_c}{V_m} - \bar{C}\frac{4V_c}{(V_m+V_c)}}\right)} \sin\left(\frac{\varphi n\pi(r - R_{max} + R_{max_t})^2}{4D\left(C_o - \varphi t \frac{V_c}{V_m} - \bar{C}\frac{V_c}{(V_m + V_c)}\right)}\right) \right] \quad (44)$$

A graph of Eq. 44 over the valid domain of $r = 0$ to $R_{max_t}$ and until $C_o$ reaches zero in media is shown in Figure 5(i-j), where the $V_m$ was again assumed to be five times that of $V_c$. Such a model is relevant not only for tissue constructs but also for bioreactor systems where cells are held in cylindrical culture constructs and perfused with a concentration gradient of nutrient.

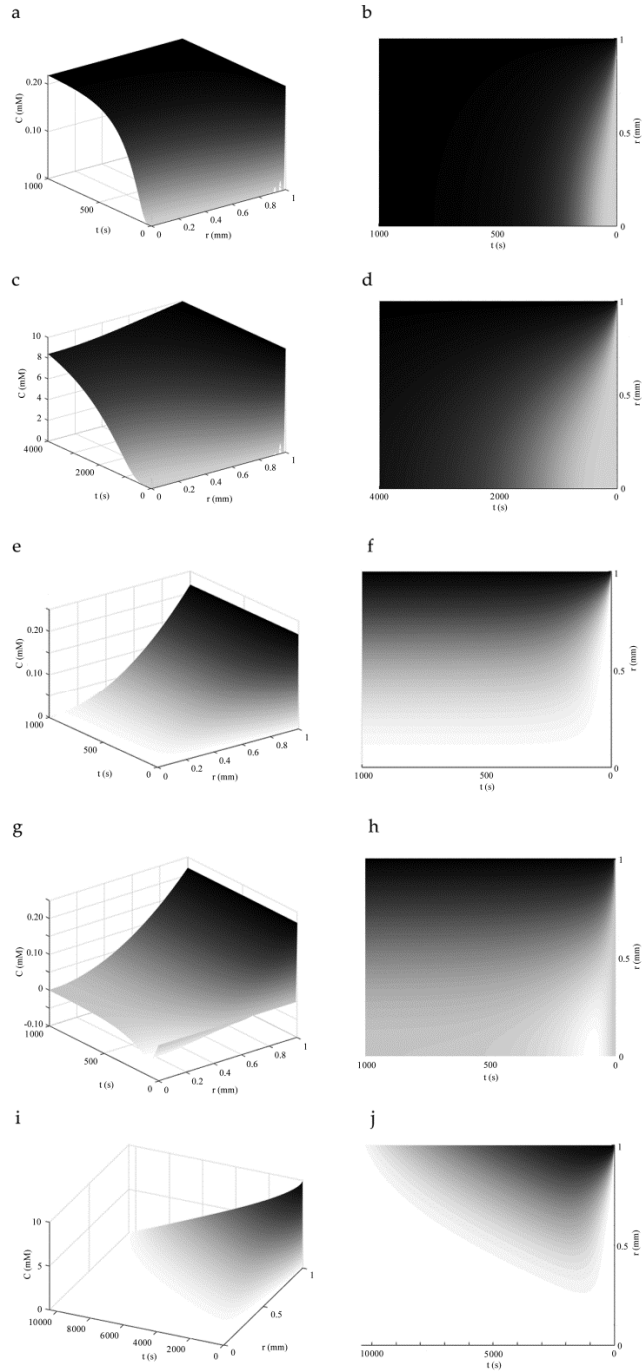

Figure 5: (a-d) Cylindrical diffusion profiles from Eq. 40 along the radial axis through time assuming no metabolism. The Bessel function solutions are shown in a cylindrical construct of 1 mm radius for the case of unlimited oxygen diffusion (a-b) and glucose diffusion (c-d); parameters for oxygen were $C_o$=0.22 mM and D=10⁻⁹ m²/s with time shown until 1000 s, and for glucose were $C_o$=10 mM and D=10⁻¹⁰ m²/s. Cylindrical diffusion profiles of oxygen diffusion in a construct with a maximized diameter of 2 mm from Eq. 43 are shown in (e-f); parameters were $C_o$=0.22 mM, $\varphi_{O_2} = 8.8x10^{-7} \frac{mol}{L \cdot s}$, and D=10⁻⁹ m²/s. The numerical solution with the same conditions as (e-f) is shown in (g-h), as explained in the text. Cylindrical diffusion profiles of limited nutrient (glucose) consumed in a construct with a maximized diameter of 2 mm described by Eq. 44 over the valid domain of $r = 0$ to $R_{max}$ and until $C_o$ reaches zero in media are shown in (i-j) assuming the media volume is only five times more than the cylindrical construct volume, with standard glucose parameters of $C_o$=10 mM, $\varphi_{Gluc} = 4.0x10^{-6} \frac{mol}{L \cdot s}$, and D=10⁻¹⁰ m²/s.

*5.1 Diffusion Models of Finite Spherical Constructs*

Spherical diffusion solutions can be formed from the three-dimensional diffusion equation by using the separation of variables method as shown in the linear and cylindrical cases. When independent functions of time and distance are substituted into the spherical diffusion equation, the following is obtained:

$$\frac{1}{DG}\frac{dG}{dt} = \frac{1}{r^2 F}\frac{d}{dr}\left(r^2\frac{dF}{dr}\right) \quad (45)$$

The general solutions produced are then $F(x) = \frac{a}{r}\cos(\varepsilon r) + \frac{b}{r}\sin(\varepsilon r)$ and $G(t) = ge^{-\varepsilon^2 Dt}$. The condition of $C(r = 0, t) = 0$ is satisfied by letting $a = 0$, meaning $F(x) = b\sin(\varepsilon r)$. Using the boundary conditions of constant nutrient at the surface of the sphere, finite nutrient levels at the center of the sphere, and no initial nutrient within the sphere as described previously, $C(R,t) = C_o$ and $C(r, 0) = 0$ for the range $0 \leq x \leq R$, a complete diffusion solution for a general spherical construct without metabolic consumption can be written as[41]:

$$C(r,t) = C_o + \frac{2RC_o}{\pi r}\left[\sum_{n=1}^{\infty}\frac{-1^n}{n}e^{-\left(\frac{n\pi}{R}\right)^2 Dt}\sin\left(\frac{n\pi r}{R}\right)\right] \quad (46)$$

This formula shows that in a non-maximized 2 mm diameter sphere with no consumption of oxygen and $D = 10^{-9}$ m²/s, it takes just less than 10 minutes for diffusion of oxygen at the center of the sphere to reach 99% of surrounding oxygen levels (Figure 6(a-d)). Likewise, for a constant ambient glucose with negligible consumption of nutrient from media (e.g., large $V_m:V_c$ ratio) and $D = 10^{-10}$ m²/s, glucose reaches 99% of surface levels at the center of a maximized sphere in just under 1.5 hours.

*5.2 Steady and Non-Steady State Model of Metabolically-Maximized Spherical Constructs with Unlimited Nutrient*

As described previously, the steady state concentration is determined by metabolism and diffusivity in the construct. The spherical equation for a maximized value of $R$ that results in zero concentration at the center of the cylinder is again implemented with boundary conditions of 1) $C(r = R_{max}, t) = C_o$, 2) $C(r = 0, t) = 0$, and 3) $C(r, t = 0) = 0$, and a complete steady and non-steady state model then becomes:

$$C(r,t) = \frac{C_o r^2}{R^2} + \frac{2C_o}{\pi}\left[\sum_{n=1}^{\infty}\frac{-1^n}{n}e^{-\left(\frac{n\pi}{R}\right)^2 Dt}\sin\left(n\pi\frac{r^2}{R^2}\right)\right] \quad (47)$$

As before, $\frac{C_o r^2}{R_{max}^2}$ simplifies to $\frac{\varphi r^2}{6D}$, as given in Eq. 23c, and, if desired, the concentration profile may be compared to the 1D case by substituting $(R_{max} - r)$ for $r$. The results of Eq. 47 are shown in Figure 6(e-f) for a 2 mm maximal diameter sphere where drops in nutrient at the surface of the sphere are negligible but nutrient is fully consumed by the center of the construct. The results for this spatial concentration profile are also supported by experimental evidence for diffusion in spherical constructs[72,79-81]. The numerical solution under the same conditions of Eq. 47 and with the same parameters used in Figure 6(e-f) is shown in Figure 6(g-h), demonstrating a transient negative nutrient concentration at an early phase where cells are presumed to continue nutrient consumption even as nutrient diffuses towards the inner region of the construct, as described in Section 3.5.

*5.3 Steady and Non-Steady State Model of Metabolically-Maximized Spherical Constructs with Limited Nutrient*

A model for limited nutrient consumed in the maximized construct can then be created as before by substituting $C_o = \left(C_o - \varphi t\frac{V_c}{V_m} - \bar{C}\frac{V_c}{(V_m + V_c)}\right)$ from Eq. 34 and $R_{max_t} = \sqrt{\frac{6D\left(C_o - \varphi t\frac{V_c}{V_m} - \bar{C}\frac{V_c}{(V_m + V_c)}\right)}{\varphi}}$ from Eq. 35 into Eq. 47 (where $\bar{C}$ is given by Eq. 49c and $R_{max_t}$ is substituted in place of $R$). As before, because these models describe constructs at a constant radius, and because the maximal depth that changes over time should be measured from the surface of the constructs, the variable $r$ in radial cases of limited nutrient should also be replaced with $(r - R_{max} + R_{max_t})$, as shown in Eq. 48.

$$C(r, t) =$$

$$\frac{\varphi(r - R_{max} + R_{max_t})^2}{6D} + \frac{2\left(C_o - \varphi t \frac{V_c}{V_m} - \bar{C}\frac{V_c}{(V_m + V_c)}\right)}{\pi}\left[\sum_{n=1}^{\infty}\frac{-1^n}{n}e^{\left(\frac{-\varphi t(n\pi)^2}{\left(6\bar{C}_o - 6\varphi t\frac{V_c}{V_m} - \bar{C}\frac{6V_c}{(V_m + V_c)}\right)}\right)}\sin\left(\frac{\varphi n\pi(r - R_{max} + R_{max_t})^2}{6D\left(C_o - \varphi t\frac{V_c}{V_m} - \bar{C}\frac{V_c}{(V_m + V_c)}\right)}\right)\right] \quad (48)$$

A graph of Eq. 48 over the valid domain of $r = 0$ to $R_{max_t}$ and until $C_o$ reaches zero in media is shown in Figure 6(i-j). A summary of all models for diffusion and metabolism of limited and unlimited nutrients in slabs, cylinders, and spheres is provided in Table 5.

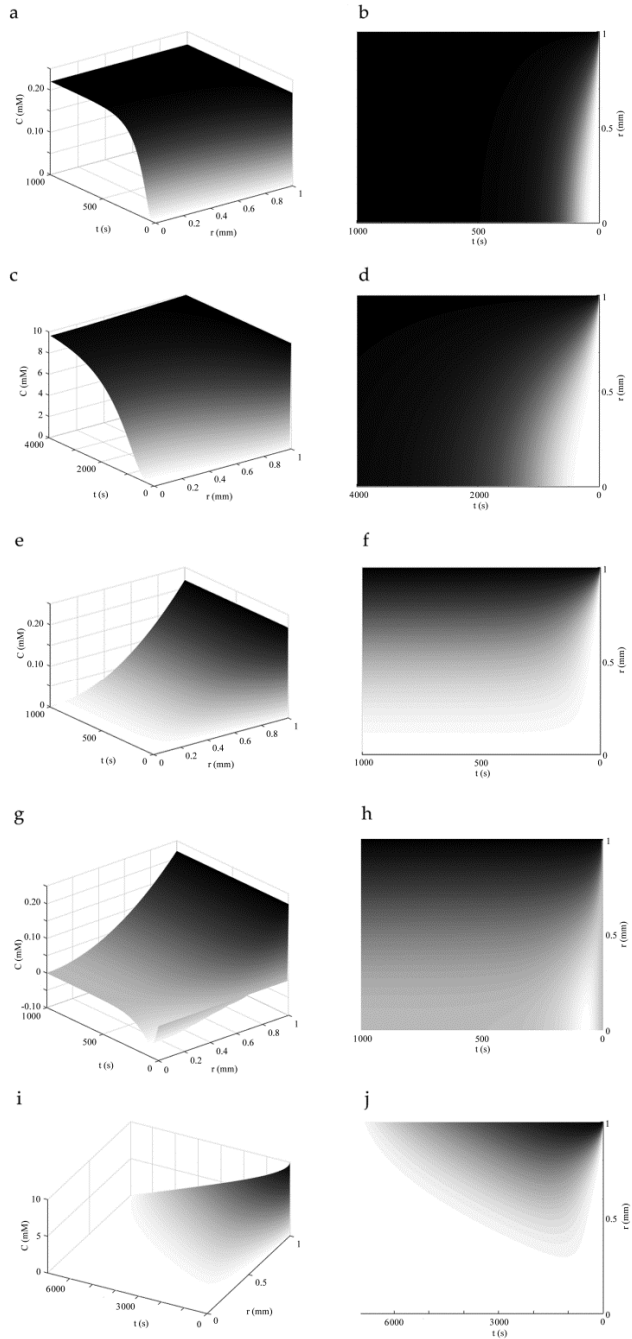

Figure 6: (a-d) Concentration profile for spherical diffusion from Eq. 46 of oxygen (a-b) using R=1 mm, $C_o$=0.22 mM, D=10⁻⁹ m²/s, and assuming no metabolism) to 1000 s, and glucose (c-d) using R=1 mm, $C_o$=10 mM, D=10⁻¹⁰ m²/s, to 4000 s. The left axis represents time and the right axis represents the radial distance into the sphere, where zero is the center of the sphere. Thus if the right axis alone were rotated about its origin in all spatial directions, it would represent the dimensions of the entire sphere. In (e-f) concentration profiles are shown for unlimited nutrient (oxygen) from Eq. 47 for a 2 mm diameter metabolically maximized sphere ($\varphi_{O_2}$ = $1.32 \times 10^{-6}\ \frac{mol}{L \cdot s}$) where nutrient is fully consumed by the center of the construct while remaining constant at the surface of the sphere. The numerical solution with the same conditions as (e-f) is shown in (g-h), as explained in the text. In (i-j) concentration profiles of a 2 mm maximal diameter sphere with limited nutrient (glucose) and zero-order metabolism described by Eq. 48 over the valid domain of $r = 0$ to $R_{max}$ and until $C_o$ reaches zero in media using $C_o$=10 mM, D=10⁻¹⁰ m²/s, $\varphi_{Gluc}$ = $6.0 \times 10^{-6}\ \frac{mol}{L \cdot s}$, and assuming the media volume is only five times greater than the spherical construct volume.

*6.1 Models of Whole-Construct Nutrient Consumption from Media Supply*

A model of the amount of an unlimited nutrient diffused into a construct can be made by integrating the molar transfer across a construct's outer surface area from time zero until the time of interest[41,45-46], as represented by the following formulas showing the material balance diffused into the construct at any point in time assuming no metabolism of nutrient:

$$\text{Slab:} \quad \bar{C} = C_o \left[ 1 - \frac{8}{\pi^2} \sum_{n=1}^{\infty} \frac{1}{(2n-1)^2} e^{-\left( \frac{(2n-1)\pi}{2T} \right)^2 Dt} \right] \quad (49a)$$

$$\text{Cylinder:} \quad \bar{C} = C_o \left[ 1 - \sum_{n=1}^{\infty} \frac{4}{(\lambda_n R)^2} e^{-\lambda_n^2 Dt} \right] \quad (49b)$$

$$\text{Sphere:} \quad \bar{C} = C_o \left[ 1 - \frac{6}{\pi^2} \sum_{n=1}^{\infty} \frac{1}{n^2} e^{-\left( \frac{n\pi}{R} \right)^2 Dt} \right] \quad (49c)$$

where $\lambda_n$ of the cylindrical equation are given by the roots of $J_o(R\lambda_n) = 0$ listed in Table 4. For limited nutrient in media, the $C_o V_m$ will drop as the nutrient diffuses into the tissue construct, with the new equilibrium concentration becoming $\frac{C_o V_m}{(V_m + V_c)}$, which can be substituted for $C_o$ above to determine final equilibrium concentration in the system. The condition of insufficient nutrient supply for a whole tissue construct, as measured from the media, can then be described by accounting for the limited nutrient consumed and diffused from ambient media. Under the conditions of constant $\varphi$, the amount of consumption by whole construct metabolism up until any time $t$ is $\varphi t V_c$. A critical threshold below which the construct's function or viability is impaired or threatened ($C_{critical}$) can be assigned to the system, which must typically be found empirically. Assuming quasi-steady-state diffusion is reached before the nutrient is completely metabolized (which is typically the case), the following formula then allows investigation of the time until critical cell dysfunction within the construct:

$$C_{media} = C_o \frac{V_m}{(V_m + V_c)} - \varphi t \frac{V_c}{V_m} \quad (50)$$

where $C_{media}$ is compared to $C_{critical}$. By setting $C_{media}$ equal to zero, the time to total consumption can also be found:

$$t_{consum} = \frac{C_o V_m^2}{\varphi(V_c V_m + V_c^2)} \quad (51)$$

Also of note, in models of maximized tissue constructs with a critical threshold of nutrient above zero, the maximal construct dimensions may simply be determined for any dimensionality $s$ by:

$$R_{max} = \sqrt{\frac{2sD(C_o - C_{critical})}{\varphi}} \quad (52)$$

*6.2 Timing and Effects of Nutrient Consumption*

Although the diffusivity of oxygen can be approximately 10 times higher than that of glucose, the available concentration of oxygen is typically one to two orders of magnitude less than glucose (glucose is typically 10-20 mM in organoid media) and the molar consumption rate of oxygen is generally higher. Thus, although glucose may take longer to reach steady state in a maximized tissue construct due to a lower diffusion coefficient, the higher concentration can still enable glucose to quickly reach adequate levels to support cell viability. Further complexity is introduced by the fact that the cell viability will depend not only on a critical threshold of nutrient concentration but also on the duration of time that the cell is deprived of the nutrient, by the fact that different cells have different thresholds of viability for different nutrients, and by the fact that one nutrient can also influence availability of other nutrients. For example, when a nutrient like glucose is low, other forms of nutrients can also be used as a compensatory energy supply, or if oxygen levels are low, anaerobic mechanisms can compensate with certain limitations. The Crabtree effect describes an increase in oxygen availability during concomitantly high glucose, or a decrease in oxygen availability during low glucose, and insufficient oxygen can likewise cause higher glucose consumption[82-84]. Therefore the determination of whether the concentration drops below a critical viability threshold must be made based on the characteristics of the cell types and the conditions of the system.

The determination of which nutrient is the limiting factor in a construct can be made by comparing thresholds and the concentration profiles between nutrients. For example, the values of $C(r,t)$ for each nutrient (either steady-state or steady- and unsteady-states for any point in time) can be subtracted from each other or graphed simultaneously and compared to determined critical thresholds. The effective $R_{max}$ can be calculated from the nutrient that gives the more limiting value, or, when individual parameters are not known, the observed $R_{max}$ for each type of nutrient can be substituted into the solutions separately for each nutrient in order to compare the concentration profiles. The models proposed above show glucose to initially be more of a temporary limiting factor for cell metabolism than oxygen due to its lower diffusivity, but after diffusion has approached steady-state, oxygen is the limiting nutrient in all cases under the given range of parameters. However, glucose limitations can also become comparable to oxygen limitations when upper limits for glucose and lower limits for oxygen are used (i.e., oxygen consumption and glucose diffusivity are minimized and glucose consumption and oxygen diffusivity are maximized).

The measured cell density in cerebral organoids was found to be $\rho = 4.82(\pm 1.41)x10^{11}\ cells/L$. With this cell density, at the upper range of oxygen diffusivity and the lower range of neural metabolic oxygen consumption, the maximal diameter of organoids is predicted to be 1.4 mm at best, while typical glucose parameters suggest a diameter of about 4 mm, and even stringent glucose parameters suggest a diameter of at least 1.4 mm, thus demonstrating how oxygen is more of a limiting nutrient than glucose at steady-state. Under the models presented herein, with $R_{max}$ dictated by oxygen and glucose consumption set in the normal range, glucose concentrations are maintained at viable levels at all depths and at all times except in the early transient stage after placing cells in hydrogel—a direct comparison of concentrations using the difference between the complete steady- and unsteady-state spherical solutions for glucose and oxygen (with glucose set at 10 mM and oxygen at 0.22 mM) shows that glucose concentration remains lower than oxygen concentration for only the first 8 min. Even though oxygen may not be as limiting as glucose in the early phase of diffusion into a tissue construct, it is plausible that the combination of low oxygen and nutrients at initiation could produce a compounding insult, threatening viability in early states of tissue formation[75]. As described, however, oxygen generally becomes the primary limiting factor at steady states in metabolically active constructs.

It is noteworthy that during periods of high neural activity, rapid spiking can increase oxygen consumption by approximately five-fold over baseline firing rates[51], and it appears that transient states of glycolysis occur over oxidative metabolism[85]. Astroglia compose about half of the number of cells in the brain[86-87], and their activity has been estimated to account for about 15% of total brain energy consumption[85]. Astrocyte metabolism may also increase during neural activity due to glutamate recycling and astrocytes may take up a significantly higher percentage of glucose and shuttle it as energy to neurons in the form of lactate[85,88]. These models do not account for metabolic coupling or metabolite production; in particular, lactic acid and carbon dioxide are metabolic products that locally increase acidity in solution, which can in turn affect material properties and cell viability, and when oxygen concentration is low or cannot meet cellular demands, lactic acid production can increase through anaerobic glycolysis. In normal neural tissue, higher neural activity is also coupled to local increases in blood flow, which serves to restore external concentration of surrounding nutrient to initial levels. Thus it may be important to account for these phenomena within a tissue construct in certain scenarios, but otherwise it can be assumed that nutrient transport systems are independent.

*7.1 Application of Diffusion Models to Cerebral Organoid Spheroids*

As described in the previous section, the maximal predicted diameter of cerebral organoids (without central cell death) is 1.4 mm. This predicted diameter is based on the physics of diffusion, the observed cell density, and a lower range of reported metabolic consumption rates for oxygen. A low metabolic activity is plausible since spontaneous neural activity appears to be infrequent in organoids, with spike activity occurring only every few minutes on average in mature organoids[31]. In addition, pluripotent cells are known to sustain quiescent states and may favor non-oxidative glycolysis over oxidative phosphorylation—it has been shown that pre-implantation mouse embryos tend to make ATP primarily by oxidative phosphorylation while the implanted embryos favor glycolysis, and similarly, naïve pluripotent stem cells tend to mix oxidative phosphorylation and glycolysis, while more primed pluripotent stem cells lean towards glycolysis[16]; other reports have suggested that glycolysis predominates in metabolic profiles of human pluripotent and hematopoietic stem cells, which may not only be a consequence of stem-like states but may also actively help mediate and maintain such states[1,2,8,15,19,89]. However, metabolic states during stages of neural differentiation are currently not well understood.

Nevertheless, even at the lowest range of reported neural oxygen consumption rates and with known parameters of cell density, diffusivity, and metabolic rate of consumption, the maximal size of the organoid was still able to exceed the predicted maximal size. Cerebral organoids grown from human induced pluripotent stem cells (hiPSCs) reached an average diameter of 1.8 mm by day 40 (Figures 7 & 8). Although many of the cerebral organoids obtained widths up to 4 mm along the longest axis, organoids typically had shorter distances between the deepest points and the nearest surface, which shortens the distance needed to perfuse the construct, and it was this shortest distance to the deepest cells that was measured. Notably, the growth characteristics of organoids closely follow those predicted by numerically-solved diffusion-limited spherical growth models[47-49], as shown in Figure 7, meaning that organoid growth is limited by diffusion or by the metabolites produced by diffusion limitations. This, along with the fact that decreased viability is known to occur at the center of organoids[31], supports the idea that diffusion depths are indeed the limiting factor in construct size.

There are several explanations for why a discrepancy in predicted versus experimental maximal diameters arose, including the idea that at least a portion of cells may be in a low metabolic state or may be using non-oxidative energy sources, meaning that $\varphi_{O_2}$ would be lower than predicted from reported neuronal metabolism values. It is also possible that cells may continue to expand the organoid in the outer regions even while cells at the center suffer ischemia. Also, because organoids are not perfect spheres, not all diffusion must occur through the measured density of metabolically active cells, or in other words, the number of cells found in the organoid over-represents how many cells would be present in a sphere of the same measured radius, and therefore the $R_{max}$ may have been underestimated by Eq. 10 due to an overestimated effective cell density. Importantly, it is also recognized that the internal structure of organoids is not homogenous, with dense cortical-like layers in outer regions and less dense progenitor regions internally. It was therefore hypothesized that regional architecture within the construct might also influence diffusion characteristics and account for observed discrepancies, and a simple model was created to investigate this question.

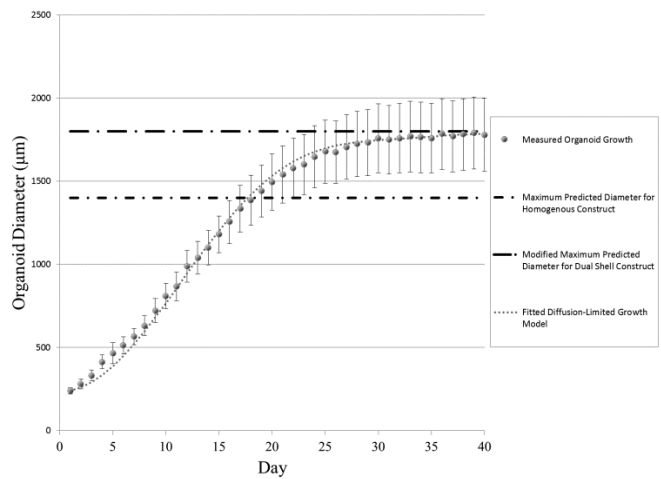

Figure 7: Graph showing the average diameter of cerebral organoids over time (±standard deviation). Organoids were able to exceed the maximal diameter predicted by typical oxygen consumption rates in homogenously dense neural constructs, but a model of dual layer cell densities in a spherical construct demonstrates a way of achieving an enhanced maximal diameter. A model of diffusion-limited growth was also applied and fitted to the data, demonstrating that growth follows expected behavior of a diffusion-limited system.

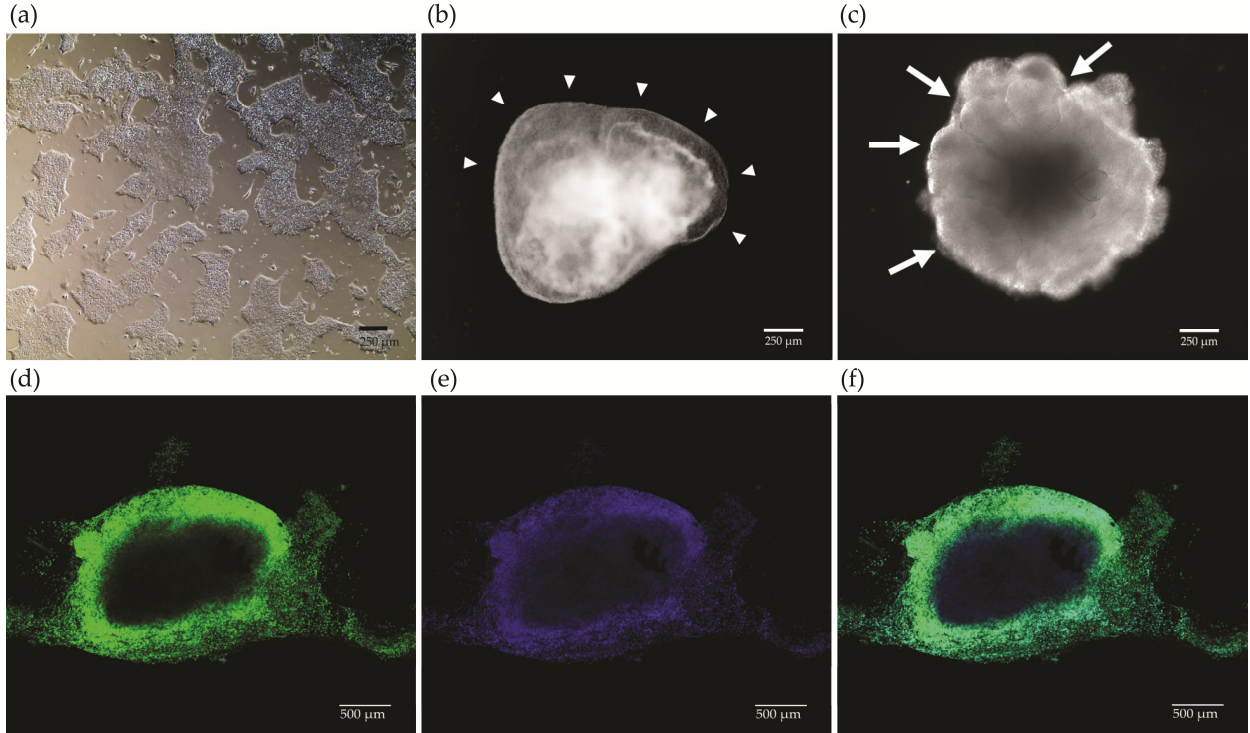

Figure 8: Induced pluripotent stem cells were cultured and passaged on a feeder-independent matrigel surface (a). Organoids demonstrated a distinct early outer layer of neuroepithelium that exhibited various forms of architecture (day 20 shown), including broad neuroepithelium (arrowheads in b) or compact localized formations of expanded neuroepithelia resembling 3D rosettes (arrows in c). In parts (d-f), stained sections of a 40-day old organoid demonstrated a dense outer region with neural identity [(d) green = TUJ1 β-III-tubulin, (e) blue = Hoechst stain, (f) merged], which according to the presented model may serve to enhance organoid size and diffusion characteristics.

### 7.2 Multi-Compartment Model and Basis for Neural Migration and Cortical Development in Avascular Spheres

An important question is whether there is an optimal architecture and regional cell density in the organoid that would maximize its potential size. To answer this question, a model was derived to represent how $R_{max}$ can change when compartmentalizing cells into an inner and outer shell of two different densities given the same amount of total cells in the construct and same nutrient diffusivity. In this model, $R_{max}$ is first set for a constant homogenous density of cells as described in Eq. 10, with any values for parameters $\varphi$, $C_o$, and $D$. The sphere is then divided into two concentric spheres, where $\Upsilon$ represents the proportional distance (from 0 to 1) where the interface between the two shells is located along the radial axis to $R_{max}$, and $\Omega$ represents the proportion of cells (from 0 to 1) within the volume of the outer shell between $\Upsilon R_{max}$ and $R_{max}$. The densities of cells in the outer and inner shells (called $\rho_2$ and $\rho_1$ respectively) can be expressed in terms of the original density ($\rho_o$), and because it is assumed that cellular metabolism rates stay the same regardless of where cells are located (although regional variations in metabolism and diffusivity could also be integrated into this model), the variable $\varphi_2$ can then be used to represent the higher consumption through the new higher density in the outer shell relative to the original $\varphi_o$, and the variable $\varphi_1$ can be used to represent the lower metabolic consumption within the inner shell relative to $\varphi_o$ (assuming, for simplicity, that the shift of cells into a new density in the outer shell occurs prior to expansion of the sphere and that the inner core can then adjust its radius based on the new level of available oxygen):

$$\varphi_2 = \frac{\Omega}{(1-\Upsilon^3)}\varphi_o \quad (53)$$

$$\varphi_1 = \frac{(1-\Omega)}{\Upsilon^3}\varphi_o \quad (54)$$

The drop in concentration across the outer shell can then be found from the following formula, where $C_s$ represents the concentration at the outer rim of the inner sphere, i.e., concentration at the interface between the two shells:

$$C_s = \frac{\varphi_2}{6D}\left((\Upsilon R_{max})^2 - R_{max}^2\right) + C_o \quad (55)$$

The new potential maximal radius of the inner shell, designated as $R_s$, can be described with the newly modified density in the inner shell, $\varphi_1$, and the new initial concentration at the outer rim of the inner shell, $C_s$:

$$R_s = \sqrt{\frac{6DC_s}{\varphi_1}} \quad (56)$$

Finally, the radial depth of the modified outer and inner shells can be summed to find a new optimized maximal radius, designated as $R_{opt}$:

$$R_{opt} = R_s + (1 - \Upsilon)R_{max} \quad (57)$$

Substituting Eq. 53 into 55, 54 into 56, 10 into 55 and 57, 55 into 56, and 56 into 57, the following relationship emerges for finding an optimized maximal radius as a function of regional architecture modifications in the sphere:

$$\frac{R_{opt}}{R_{max}} = (1 - \Upsilon) + \sqrt{\left(\frac{\Upsilon^3}{1-\Omega}\right)\left(1 - \frac{\Omega(\Upsilon+1)}{\Upsilon^2 + \Upsilon + 1}\right)} \quad (58)$$

This solution is graphed in Figure 9 and shows that when as many of the cells as possible migrate into as thin of outer shell as possible, the sphere can obtain a maximal radius several times greater than if the same number of cells were homogenously distributed throughout the sphere. As an example, if 90% of cells ($\Omega = 0.9$) in a homogenously cellularized sphere are moved into the outer half-volume of the sphere (defined at $r = R/\sqrt[3]{2}$ or $\Upsilon$=0.79), then the new achievable maximum radius ($R_{opt}$) is 1.5 times greater than the original $R_{max}$. In these cerebral organoids, the observed maximal diameter of 1.8 mm (which is 1.3 times greater than the 1.4 mm limit suggested by the homogenous cellular density model) would suggest that approximately 85% of cells ($\Omega$ =0.85) must be in the outer half-volume of the sphere ($\Upsilon$=0.79), as opposed to the 50% that would be expected in the same volume in a homogenous construct, which matches well with the observation that neural precursor cells migrate outwardly in early stages to form high density cortical-like structures in both brain and organoids[31,90-91] (also see Figure 8). Clearly this is only a simplified model that neglects the fact that as regions shift and expand, concentration profiles, shell thicknesses, and cell densities will change in response to each other, and mechanical limitations prohibit the formation of an infinitely thin outer shell of cells, but this model demonstrates the concept that diffusion through a thin but dense layer of larger surface area is more efficient than diffusion through a sparse but thick layer of smaller surface area.

It is often suggested that a larger and more convoluted cortex confers the advantage of a larger surface area for cortical networks, but it is not explained why neurons must be on the outer surface in the first place. The above model is an algebraic demonstration showing how the largest potential size of a cellularized sphere is achieved by localizing the majority of metabolically active neurons into a thin and dense outer shell, and it may hold many implications for why inchoate and avascular neural systems develop with neural precursors migrating outwardly. A multi-compartment model may serve to illustrate internal architecture and subspecialized functions of cells within cerebral organoids, as shown in Figure 9. At the highest oxygen concentrations, metabolically-active cells like neurons can best function in a dense outer zone (zone 1) near a high oxygen supply, and this external localization also best enables growth of the organoid and corresponding expansion of neurons in this region. In the second zone, marginally lower oxygen levels may promote expansion of neuroglial precursor populations, a phenomenon which has been observed in cell cultures with controlled hypoxia levels[11-13]. In the central region of still lower oxygen levels, stem cell pluripotency may be maintained, as has also been observed in cell cultures with controlled hypoxia levels[4-6,14], although other evidence has also suggested that hypoxia can induce differentiation[9-10,18]. Finally, zone 4 represents a potential hypoxic region where cell function and viability are threatened.

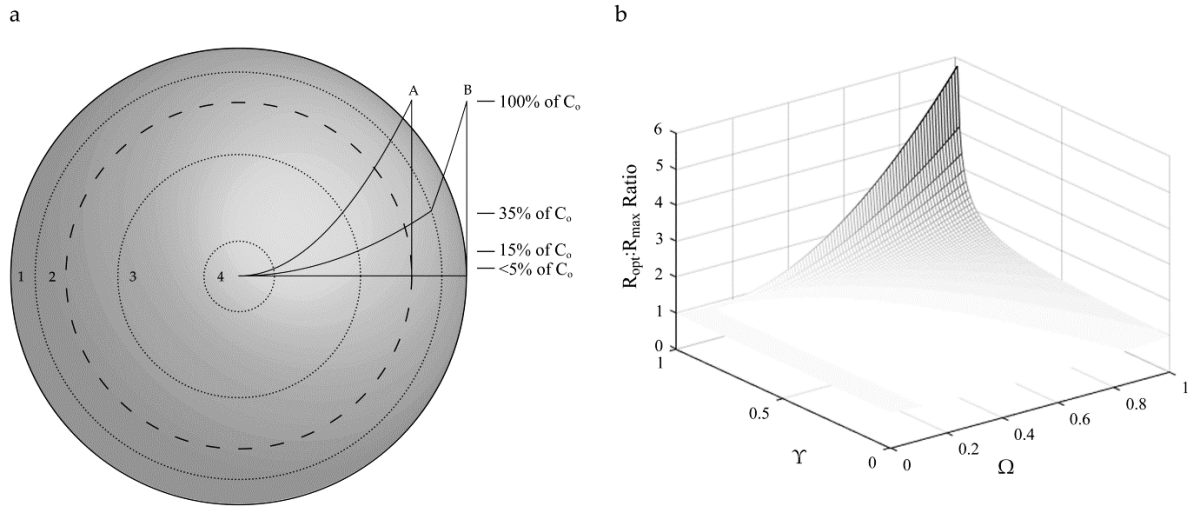

Figure 9: A multi-compartment spherical model for cerebral organoids is shown in part (a). Curve A demonstrates the general concentration profile for any metabolized gas or nutrient and the maximal diameter of 1.4 mm (dashed line) for a homogenous distribution of cells, while curve B demonstrates the concentration profile and modified maximal radius when a percentage of cells shift into a denser outer rim (zone 1), all other parameters remaining constant, thereby enabling an enhanced maximal diameter of 1.8 mm (solid border). The percentage of ambient nutrient concentration within the construct is also marked on the concentration profile curves. The maximal diameter is enhanced when more metabolically active cells localize densely into zone 1. Zone 2 may represent an intermediate region of progenitor cells, zone 3 may represent a region of multipotency preserved by hypoxic conditions, and zone 4 represents a potentially ischemic region. Part (b) is a graph of Eq. 58 demonstrating how the maximal achievable radius of a spherical construct increases when a homogenously cellularized sphere is modified into a two shell sphere with a higher percentage of the cells ($\Omega$) placed into the outer shell beginning at point $\Upsilon$ along the radius between 0 and $R_{max}$.

# Discussion

The creation and implementation of functional engineered tissue constructs requires overcoming numerous challenges, one of the greatest of which in both *in vitro* constructs and *in vivo* implants involves diffusion limitations that may dictate size constraints and affect cell function and viability. This work therefore illustrates new theoretical models that are broadly applicable to numerous problems and experiments in tissue engineering, allowing researchers to infer diffusion and metabolism processes within tissue constructs based on fundamental principles of physics. Steady and non-steady state models are provided for certain 3D construct designs, including slabs, cylinders, and spheres, the results of which are otherwise generally not obtainable except through numerical approximation methods. The solutions described in this work will help enable modeling of oxygen and nutrient delivery to cells, understanding of molecular mass transport signaling in cell function, study of cell viability in 3D constructs, control of stem cell states or states of decreased nutrient delivery, and design of 3D constructs with improved diffusion capabilities, such as by choice of architecture and materials, as well as additional design features such as fluid vents or perfusion channels. This work is therefore important in the study of development and disease models, and may also be relevant for conditioning cells to survive after insults like ischemia or implantation into the body[92].

The presented models demonstrate how parameters for various states of cell growth and metabolism can be used to describe complex diffusion behaviors and nutrient concentrations in great detail. Although the equations presented do not provide solutions for certain conditions that change in time, such as proliferation of cells, growth of a construct, or variations in metabolic rates through time, each state can be determined given any conditions at any point in time. In addition, many of the methods and approaches used herein can be applied to more complex solutions for construct models in three-dimensional coordinate systems. Models of multilayered effects that require numerical simulation have been described in the case of tumor growth[47-49], for example, which may also serve as a basis for more complex models of regionally-distinct internal compartments and processes in tissue constructs.

Growth characteristics of cerebral organoids correlate with mathematical models of diffusion-limited growth, and models using known diffusion and metabolism parameters show oxygen to be a major limiting factor within organoids. It was also found that the maximal spherical radius predicted by known parameters and general homogenous spherical models may in fact be exceeded in culture at least in part by localizing a high density of metabolically active cells into the outer layers of the organoid. The formation of radially-aligned cells and internal fluid chambers may also further enhance diffusion and convection within neural tissue constructs. Thus the architecture of the brain may be structured at least in part based on underlying physical principles. It is not yet known to what extent cortical differentiation, regionalization, layer formation, and anatomical structure develop as functions of diffusion processes that affect concentrations of nutrient and signaling molecules, but physical diffusion characteristics may influence many endogenous signaling pathways, and it is at least plausible that the migration of radial glia and neural progenitor type cells towards a dense collection in the outer cortical layers of the brain, as is observed both in organoids and *in vivo*[31,90-91], not only imparts an advantage of expansive surface area for large-scale neural networks but also is optimized within physical constraints to best allow diffusion of nutrients and metabolites to and from these highly energetic regions with minimal diffusion distance, minimal vascularization impeding the formation of dense neural networks, and circulating cerebrospinal fluid that can help maintain nutrient concentrations at the inner and outer surfaces of neural tissue.

This work provides a framework for better understanding characteristics and behaviors of tissue constructs, and although the accuracy of these theoretical models may vary depending on the actual conditions and parameters of the system, experimental results have proven to be a consistent with the underlying physics of diffusion in many experimental designs[43-44,57-64,72,75,79-81,93-94]. Future work will expand the set of solutions for a variety of shapes and coordinate systems, as well as for additional orders of metabolism, in order to comprehensively cover the variety of 3D tissue constructs that are relevant to tissue engineering and to understand the operation and design of biological phenomena based on physical principles. The continued investigation of 3D engineered tissue will be essential to successfully understanding tissue models and for implementing engineered tissue constructs for clinical use.

# Tables

| Table 1: Typical Concentrations and Coefficients in Tissue Diffusion Systems | |
|---|---|
| Parameter: | Typical Values for Diffusion Coefficients ($D$) and Initial Concentrations ($C_o$): |
| Diffusion Coefficient ($D$) in Tissue or Hydrogel | $1x10^{-10}$ to $2.5x10^{-9} \frac{m^2}{s}$ [43-44,51-64,95-97] |
| $C_o$ of Oxygen in Liquid Media | 0.20 to 0.22 mM [44,57,65,68] |
| $C_o$ of Oxygen in Arterial Blood† | 5.5 to 10 mM [98-99] |
| $C_o$ of Oxygen in Venous Blood† | 3 to 7.5 mM [98-99] |
| $C_o$ of Glucose in Culture Media | 4 to 55 mM (typically 5 to 20 mM) |
| $C_o$ of Glucose in Normal Blood | 5 to 6 mM |
| $C_o$ of Glucose in Normal CSF | 3 to 4.5 mM |

Table 1: Summary of common ranges for the diffusion coefficient and baseline concentration in various solutions and in typical tissue and culture conditions. †It is helpful to note that blood oxygen values are based on normal hemoglobin levels, that less than 2% of oxygen in blood is freely dissolved unbound to hemoglobin, that oxygen dissociation from hemoglobin follows nonlinear kinetics, and that normal jugular venous saturation from cerebral blood flow can be lower (55-75% $S_vO2$) than normal mixed venous saturation.

| Table 2: Typical Oxygen Consumption Rates | |
|---|---|
| Cell Type: | Metabolic Approximation for Oxygen Consumption ($m_{o_2}$): |
| Average Neuron of the Human Brain | $7.7x10^{-16} \frac{mol}{cell \cdot s}$ [69,100] |
| Average Neuron of the Brain Across Species | $1.1x10^{-17}$ to $1.0x10^{-15} \frac{mol}{cell \cdot s}$ [69-71] |
| Human Embryonic Stem Cell (ESC)‡ | $8.3x10^{-19}$ to $1.5x10^{-16} \frac{mol}{cell \cdot s}$ [15,19, 101-102] |
| Human Induced Pluripotent Stem Cell (iPSC)‡ | $1.8x10^{-18}$ to $2.3 x10^{-18} \frac{mol}{cell \cdot s}$ [19] |
| ESC (Rodent) ‡ | $3.0x10^{-17}$ to $4.0x10^{-17} \frac{mol}{cell \cdot s}$ [103] |
| Human Mesenchymal Stem Cell (MSC)‡ | $2.0x10^{-17}$ to $3.8x10^{-17} \frac{mol}{cell \cdot s}$ [104] |
| Chondrocyte (Bovine) | $2.0x10^{-19}$ to $6.0x10^{-18} \frac{mol}{cell \cdot s}$ [43,44,62] |
| Human Hepatocyte | $1.0x10^{-16}$ to $9.0x10^{-16} \frac{mol}{cell \cdot s}$ [105-107] |
| Hepatocyte (Rodent) | $2.0x10^{-16}$ to $9.0x10^{-16} \frac{mol}{cell \cdot s}$ [105,108] |
| Human Cardiomyocyte (including iPSC-derived) | $3.3x10^{-17}$ to $2.2 x10^{-16} \frac{mol}{cell \cdot s}$ [109-110] |
| Cardiomyocyte (Rodent) | $2.5x10^{-17}$ to $1.7x10^{-16} \frac{mol}{cell \cdot s}$ [105,111-117] |
| Renal Epithelial Cells (Dog) | $2.8x10^{-17} \frac{mol}{cell \cdot s}$ [105] |
| Human Fibroblast (including ESC-derived) | $8.3x10^{-19}$ to $1.8x10^{-17} \frac{mol}{cell \cdot s}$ [19,105] |
| Vascular Endothelial Cell (Bovine) | $2.8x10^{-17}$ to $5.0x10^{-17} \frac{mol}{cell \cdot s}$ [113] |
| Average Cell of the Human Body | $2.5x10^{-18} \frac{mol}{cell \cdot s}$ [105] |
| Various Cancer Cell Types | $3.7x10^{-18}$ to $5.5x10^{-15} \frac{mol}{cell \cdot s}$ [81,105] |

Table 2: Summary of metabolic ranges of oxygen consumption for various cell types as available in the literature. The metabolic rate ($m$) can be used to calculate the metabolic consumption of the construct by the methods described in the text. Pluripotent stem cells tend to have low metabolic rates, while progenitor cells and differentiated cells have higher metabolic rates. Neurons that are spontaneously active will have higher metabolic ranges while neurons that are not actively firing will tend to be in the lower ranges. The metabolic rate of cortical neurons is significantly higher than that of cerebellar neurons, but cerebellar neurons make up a slightly larger percentage of the total number of neurons in the brain[69,118]. These are only values reported under certain conditions, and actual values may vary depending on the state and environment of the cells. Values have been converted from the original citations where necessary to make the units consistent.

| Table 3: Typical Glucose Consumption Rates | |
|---|---|
| Cell Type: | Metabolic Approximation for Glucose Consumption ($m_{Gluc}$): |
| Average Neuron of the Human Brain | $9.1x10^{-17} \frac{mol}{cell \cdot s}$ [69,100] |
| Average Neuron of the Brain Across Species | $8.1x10^{-17}$ to $1.2x10^{-16} \frac{mol}{cell \cdot s}$ [69] |
| Average Cortical Neuron of the Brain Across Species | $1.9x10^{-16}$ to $3.9x10^{-16} \frac{mol}{cell \cdot s}$ [69] |
| Average Cerebellar Neuron of the Brain Across Species | $9.3x10^{-18}$ to $2.2x10^{-17} \frac{mol}{cell \cdot s}$ [69] |
| Average Cortical Neuron of Human | $2.2x10^{-16} \frac{mol}{cell \cdot s}$ [69] |
| Average Cerebellar Neuron of Human | $1.1x10^{-17} \frac{mol}{cell \cdot s}$ [69] |
| Human Mesenchymal Stem Cell (MSC)‡ | $3.5x10^{-17}$ to $1.4x10^{-16} \frac{mol}{cell \cdot s}$ [104,119] |
| Chondrocyte (Bovine) | $3.7x10^{-17} \frac{mol}{cell \cdot s}$ [43] |

Table 3: Summary of metabolic ranges of glucose consumption for various cell types as available in the literature. These are only values reported under certain conditions, and actual values may vary depending on the state and environment of the cells. Other sources of energy besides those listed here may also be used simultaneously by cells, particularly lipids, pyruvate, lactate, amino acids, or stored glycogen. Values have been converted from the original citations where necessary to make the units consistent. ‡States of glycolysis and oxidative phosphorylation may be highly correlated with states of pluripotency or differentiation, as described in the text, and cell state may therefore affect these values significantly.

| | $J_o(x)$ ($n^{th}$ roots) | $\lambda_n$ for $J_o(\lambda_n R)$ |
|---|---|---|
| n=1 | 2.4048 | 2.4048/R |
| n=2 | 5.5201 | 5.5201/R |
| n=3 | 8.6537 | 8.6537/R |
| n=4 | 11.7915 | 11.7915/R |
| n=5 | 14.9309 | 14.9309/R |
| n=6 | 18.0711 | 18.0711/R |
| n=7 | 21.2116 | 21.2116/R |
| n=8 | 24.3525 | 24.3525/R |
| n=9 | 27.4935 | 27.4935/R |
| n=10 | 30.6346 | 30.6346/R |

Table 4: The first ten roots of the first kind of Bessel function of order zero.

| Construct Design | Maximum Depth | Complete Transient and Steady-State Model of Construct in Unlimited Nutrient | Complete Transient and Steady-State Model of Construct in Limited Nutrient |
|---|---|---|---|
| 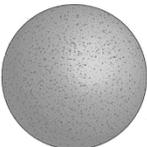 | Eq. (10) | Eq. (47) | Eq. (48) |
| 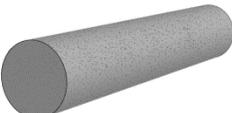 | Eq. (13) | Eq. (43) | Eq. (44) |
| 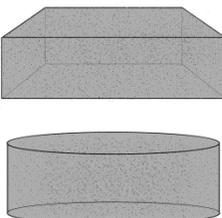 | Eq. (4) | Eq. (37) | Eq. (38) |

Table 5: Summary of important equations described for tissue construct modeling.

# Acknowledgements

I would like to thank Dr. Alex Shcheglovitov for his kind support and providing the iPSC line.

# Disclosure Statement

The author affirms that no competing financial interests exist. No grant funding was provided for this work.

# Appendix

The diffusion equation can be solved numerically in MATLAB using the PDE solver function for boundary value problems or by implementing other finite difference schemes. The following describes how to implement basic numerical solutions in homogeneously cellularized constructs using the MATLAB *pdepe* function. Four MATLAB script files will need to be created, which can be named as follows: 1) PDE_Diffusion_Solution.m, 2) pdex1bc.m, 3) pdex1ic.m, and 4) pdex1pde.m. The first file will call on the others to find the solution. By setting parameters for the thickness of the construct, the time duration, initial concentration in the construct and at the surface, diffusion coefficient, cellular metabolic rate, and cell density, the diffusion of oxygen into planar, cylindrical, or spherical constructs can be found:

```
T=4; %Tmax (in mm), or Rmax for radial conditions
t_end=1000; %End time point (in s)
Co=0.00022; %(in mol/L)
x = linspace(0,T,1000);
t = linspace(0,t_end,1000);
D=0.001; %(in mm^2/s))
ρ=3.5*10^8; %cell density (in cell/L)
φ=(8.1*10^-17).*rho; %metabolic rate (in mol/Ls)
```

The solution is then found from the following function:
```
C = pdepe(s,@pdex1pde,@pdex1ic,@pdex1bc,x,t);
```

where s=0 for the unitary spatial dimension (no radial symmetry), or s=1 for cylindrical symmetry, or s=2 for spherical symmetry. The solution can then be graphed as follows:
```
C = C(:,:,:);
surf(x,t,C);
xlabel('x');ylabel('t');zlabel('C');
```

pdex1bc:
```
function [pl,ql,pr,qr] = pdex1bc(xl,Cl,xr,Cr,t)
pl = Cl-Co; %left Dirichlet boundary condition (Co for linear condition)
ql = 0; %left Neumann boundary condition
pr = Cr-0; %right Dirichlet boundary condition (Co for radial conditions)
qr = 0; %right Neumann boundary condition
```

pdex1ic:
```
function u0 = pdex1ic(x)
u0 = 0; %At time=0 the initial concentration in the construct (Cᵢ) is zero.
```

pdex1pde:
```
function [c,f,φ] = pdex1pde(x,t,u,DuDx)
c = 1;
f = D.*DuDx; %Insert the value for D.
φ = 0; %Insert value used for Tmax or Rmax (negative indicates consumption).
```

The analytic solutions presented in this paper may be plotted with the following MATLAB scripts, with the metabolically-maximized 1D slab solutions shown here as examples.

1D Slab with Metabolism of Unlimited Nutrient:
```
D=0.001; %in mm^2/s
Co=0.00022; %in mol/L
φ=2.75e-08; %in mol/Ls
T=4; %in mm
t_end=4000; %in s
x = linspace(0,T,100);
t = linspace(0,t_end,100);
[x,t]=meshgrid(x,t);
```

```
Z=0;
i=1000;
for n=1:i;
    Z=Z+((((1./n).*exp(-D.*t.*(n.*pi./T).^2)).*(sin(n.*pi.*(((2.*φ.*T.*x-
        (φ.*x.^2)))./(2.*Co.*D))))));
end
C=((φ./(2.*D)).*x.^2- φ.*T.*x./D)+Co-(2.*Co./pi).*Z;
surf(x,t,C);
xlabel('x');
ylabel('t');
zlabel('C');
axis([0 T 0 t_end 0 Co]);
```

### 1D Slab with Metabolism of Limited Nutrient:

```
D=0.0001; %in mm^2/s
C_o=0.01; %in mol/L
φ=1.25e-07; %in mol/Ls
T_max=4; %in mm
t_end=50000; %in s
x = linspace(0,T_max,100);
t = linspace(0,t_end,100);
[x,t]=meshgrid(x,t);
V_c=1;
V_m=5;
Z2=0;
i=1000;
for n=1:i;
    Z2=Z2+(1./((2.*n-1).^2)).*exp(-((((2.*n-1).*pi)./(2.*T_max)).^2).*D.*t);
end
C_bar=C_o.*(1-(8./(pi.^2)).*Z2);
T=sqrt(2.*(C_o-φ.*t.*(V_c./V_m)-C_bar.*(V_c./(V_c+V_m))).*D./φ);
x(x(:,:)>T(:,:))=nan;
Z1=0;
i=1000;
for n=1:i;
    Z1=Z1+((((1./n).*exp(-D.*t.*(n.*pi./T).^2)).*(sin(n.*pi.*(((2.*φ.*T.*x-
        ((φ).*x.^2)))./(2.*C_o.*D))))));
end
C=((φ./(2.*D)).*x.^2-φ.*T.*x./D)+(C_o-φ.*t.*(V_c./V_m)-C_bar.*(V_c/(V_c+V_m)))-
    ((2.*(C_o-φ.*t.*(V_c./V_m)-C_bar.*(V_c/(V_c+V_m))))./pi).*Z1;
surf(x,t,C);
xlabel('x');
ylabel('t');
zlabel('C');
axis([0 T_max 0 t_end 0 C_o]);
```